\pdfoutput=1

\documentclass[aps,prl,longbibliography,reprint,amssymb,amsmath,floatfix]{revtex4-1}

\usepackage[utf8]{inputenc}
\usepackage[T1]{fontenc}

\usepackage{newtxtext}
\usepackage{textcomp}
\usepackage{newtxmath}
\usepackage{bm}
\usepackage{amsmath}

\usepackage[english]{babel}
\usepackage{csquotes}

\usepackage{microtype}

\usepackage{graphicx}

\usepackage{pgf}
\usepackage{tikz}

\usepackage{siunitx}
\usepackage[version=3]{mhchem}

\usepackage{xcolor}
\definecolor{DarkRed}{rgb}{0.65,0,0}
\definecolor{DarkBlue}{rgb}{0,0,0.65}
\definecolor{DarkGreen}{rgb}{0,0.65,0}
\definecolor{DarkYellow}{rgb}{0,0.65,0.65}

\usepackage[colorlinks=true,allcolors=blue]{hyperref}
\addto\extrasenglish{\def\equationautorefname~#1\null{Eq.~(#1)\null}}
\AtBeginDocument{

}

\renewcommand{\vec}[1]{\bm{#1}}
\providecommand{\drm}{\ensuremath{\mathrm{d}}}

\providecommand{\sym}[1]{\mathbf{#1}}
\providecommand{\symsub}[2]{\mathbf{#1}_{\bm{#2}}}
\providecommand{\abs}[1]{\lvert#1\rvert} 

\providecommand{\ie}{\textit{i.e.}}
\providecommand{\eg}{\textit{e.g.}}
\providecommand{\prlsection}[1]{\textit{#1}.\kern.05em---\kern.05em\ignorespaces}

\begin{document}

\title{Current Control of Magnetism in Two-Dimensional \ce{Fe3GeTe2}}
\author{Øyvind Johansen}
\email{oyvinjoh@ntnu.no}
\author{Vetle Risinggård}
\email{vetle.k.risinggard@ntnu.no}
\author{Asle Sudbø}
\author{Jacob Linder}
\author{Arne Brataas}
\affiliation{
  Center for Quantum Spintronics, 
  Department of Physics, 
  Norwegian University of Science and Technology, 
  N-7491 Trondheim, 
  Norway}
\date{\today}

\begin{abstract}
 The recent discovery of magnetism in two-dimensional van der Waals systems opens the door to discovering exciting physics.
 We investigate how a current can control the ferromagnetic properties of such materials. Using symmetry arguments, we identify a recently realized system in which the current-induced spin torque is particularly simple and powerful. In \ce{Fe3GeTe2}, a single parameter determines the strength of the spin--orbit torque for a uniform magnetization. The spin--orbit torque acts as an effective out-of-equilibrium free energy.
 The contribution of the spin--orbit torque to the effective free energy introduces new in-plane magnetic anisotropies to the system.
 Therefore, we can tune the system from an easy-axis ferromagnet via an easy-plane ferromagnet to another easy-axis ferromagnet with increasing current density. This finding enables unprecedented control and provides the possibility to study the Berezinski{\v i}--Kosterlitz--Thouless phase transition in the 2D $XY$ model and its associated critical exponents. 
\end{abstract}

\maketitle

\prlsection{Introduction}
Magnetism in lower dimensions hosts interesting physics that has been studied theoretically for many decades.
Examples include the intriguing physics of the exactly solvable 2D Ising model~\cite{Onsager} and the Berezinski{\v i}--Kosterlitz--Thouless (BKT) phase transition in the 2D $XY$ model~\cite{Berezinskii1,Berezinskii2,KosterlitzThouless}.
However, experimentally realizing the details of these theoretical predictions has proven difficult. One reason for this difficulty is that fabricating atomically thin films is challenging. The isolation of graphene in 2004 provided a path for exploring two-dimensional van der Waals materials~\cite{Novoselov}.
Creating two-dimensional films that have long-range magnetic order at finite temperatures is more challenging because of the Mermin--Wagner theorem~\cite{MerminWagner}.
This theorem states that long-range magnetic order does not exist at finite temperatures below three dimensions when the exchange interaction has a finite range and the material has a continuous symmetry in spin space.
Consequently, realizing two-dimensional magnetic materials requires breaking the continuous symmetry of the system, \eg\, by a uniaxial magnetocrystalline anisotropy.
This provides an energy cost (also known as a magnon gap) to suppress long-range fluctuations that can destroy the magnetic order.
The recent discovery of magnetic order in two-dimensional van der Waals materials has therefore led to a large number of studies of magnetism in atomically thin films~\cite{Burch2018}.
Magnetic order has been reported in \ce{FePS3}~\cite{Lee2016},
\ce{Cr2GeTe6}~\cite{Gong2017},
\ce{CrI3}~\cite{Huang2017},
\ce{VSe2}~\cite{Bonilla2018},
\ce{MSe_{$x$}}~\cite{OHara},
and \ce{Fe3GeTe2}~\cite{Fei2018,Deng2018a}.
In addition, multiferroicity has been identified in \ce{CuCrP2S6}~\cite{Lai}.
These new two-dimensional magnets are amenable to electrical control~\cite{Deng2018a,Huang,Jiang,Wang} and produce record-high tunnel magnetoresistances~\cite{Kim}.

Currents can induce torques in magnetic materials~\cite{Brataas2012}. In ferromagnets with broken inversion symmetry, the spin--orbit interaction leads to spin--orbit torques (SOTs)~\cite{Manchon2008}. These torques can be present even in the bulk of the materials without requiring additional spin-polarizing elements. The effects of SOTs are typically sufficiently large to induce magnetization switching or motion of magnetic textures~\cite{Manchon}. With the rich physics that is known to exist in two-dimensional magnetic systems, we explore how currents can provide additional control over the magnetic state via SOTs. 

Although many of the newly discovered two-dimensional magnetic systems exhibit SOTs, we find that in one material the torque is particularly simple and powerful. The form of the torque is simple because it is determined by a single parameter. The torque is also influential in determining the magnetic state of the system. In contrast to many other systems, we can describe the current-induced effects via an effective out-of-equilibrium free energy. Therefore, the SOT enables unprecedented control over the magnetic state via the current. We will demonstrate how the current can drive the system from having easy-axis anisotropy along one direction to anisotropy along a different axis by proceeding via an intermediate state with easy-plane anisotropy. 

Interestingly, the current-induced easy-plane configuration provides the possibility to study the BKT phase transition in this system. The BKT transition is an example of a so-called conformal phase transition in which the scale invariance of a topologically ordered state, \ie\, conformal invariance,  is lost at the (topological) phase transition~\cite{Kaplan2009conformality}. When driven by a current, we realize a 2D conformal field theory in the low-temperature phase, with conformality being lost~\cite{Kaplan2009conformality} at the transition to the paramagnetic phase.  Additionally, it was recently discovered that an ionic gate considerably increases the critical temperature~\cite{Deng2018a}. Consequently, two-dimensional \ce{Fe3GeTe2} forms an ideal and very rich laboratory for studying fundamental problems of broad current interest in condensed  matter physics and beyond at elevated temperatures.

\begin{figure}[t]
  \includegraphics[width=\columnwidth]{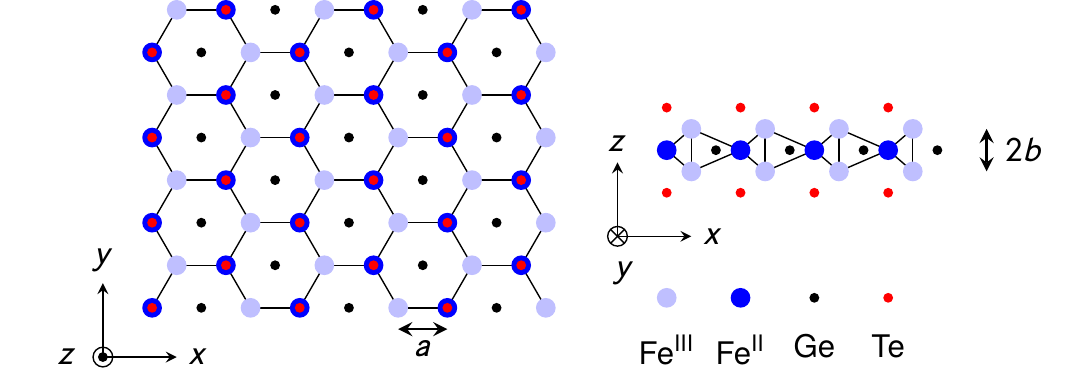}
  \caption{\label{fig:crystal}
  Crystal structure of an \ce{Fe3GeTe2} monolayer. 
	Left: view along $\vec e_z$; 
	right: view along $\vec e_y$. 
	$a$ is the in-plane bond length between \ce{Fe^{III}} and \ce{Fe^{II}}.
	$2b$ is the out-of-plane distance between the two \ce{Fe^{III}} sublattices.
	\ce{Fe^{III}} and \ce{Fe^{II}} represent the two inequivalent \ce{Fe} sites in oxidation states $+3$ and $+2$, respectively. 
	Redrawn after Ref.~\cite{Deng2018a}.
  }
\end{figure}

\prlsection{System}
We consider a monolayer of \ce{Fe3GeTe2}. \autoref{fig:crystal} shows the crystal structure of this material. \ce{Fe3GeTe2} crystallizes in the hexagonal system, space group 194, point group $\sym 6/\sym m\,\sym 2/\sym m\,\sym 2/\sym m$, known as $D_{6h}$ in the Schönflies notation~\cite{Deiseroth2006}. However, the basis reduces the point group symmetry to $\sym{\bar 6m2}$ ($D_{3h}$). Placing an \ce{Fe3GeTe2} monolayer on a substrate may reduce the symmetry even further (point group $\sym{3m}$) if the bottom tellurium layer hybridizes with the surface. Here, we assume that a possible monolayer--substrate interaction is weak.

The SOT can be written as~\cite{Hals2013,*Hals2015}
\begin{equation}\label{eq:torque}
\vec\tau=-\abs{\gamma}\vec m\times\vec H_\text{SOT},
\end{equation}
where $\gamma$ is the gyromagnetic ratio and $\vec m$ is the magnetization unit vector. For a spatially uniform magnetization, the effective magnetic field $\vec H_\text{SOT}$ due to the SOT in an \ce{Fe3GeTe2} monolayer is~\footnote{
	See the Supplemental Material, which includes Ref.~\cite{Birss1964}, a derivation of the spin--orbit torques, and the details of the critical-temperature calculation.
}
\begin{equation}\label{eq:heff}
\vec H_\text{SOT}=\Gamma_0[(m_xJ_x-m_yJ_y)\vec e_x-(m_yJ_x+m_xJ_y)\vec e_y].
\end{equation}
Here, $m_i$ are magnetization components, and $J_i$ are components of the current density. 
$\Gamma_0$ is a free parameter that is determined by the spin--orbit coupling. 

We provide a rigorous derivation of the effective field $\vec H_\text{SOT}$ based on Neumann's principle in the Supplementary Material~\cite{Note1}. 
In \ce{Fe3GeTe2}, we can understand the dependence of the SOT on the magnetization and currents in \autoref{eq:heff} as follows. 
The crystal structure in \autoref{fig:crystal} is invariant under 
a three-fold rotation about the $z$ axis ($\symsub{3}{z}$),
an inversion of the $y$ axis ($\symsub{m}{y}$),
and an inversion of the $z$ axis ($\symsub{m}{z}$). 
These symmetry operations generate the point group $\sym{\bar 6m2}$.
Since $\vec H_\text{SOT}$ only contains terms that are quadratic in $y$, it is invariant under the operation $\symsub{m}{y}$.
The operation $\symsub{3}{z}$ transforms $(m_x,m_y)$ into
\begin{equation} \label{eq:3ztrans}
\frac{1}{2}\left(\begin{array}{rr} -1 & \sqrt{3} \\ -\sqrt{3} & -1 \end{array}\right)\left(\begin{array}{cc} m_x \\ m_y \end{array}\right)=\frac{1}{2}\left(\begin{array}{c} -\phantom{\sqrt{3}}m_x+\sqrt{3}m_y \\ -\sqrt{3}m_x-\phantom{\sqrt{3}}m_y \end{array}\right),
\end{equation}
and similarly for $(J_x,J_y)$ and $(\vec e_x,\vec e_y)$. 
Back-substitution of the transformation in \autoref{eq:3ztrans} into \autoref{eq:heff} shows that $\vec H_\text{SOT}$ is also invariant under this operation.
The effective field $\vec H_\text{SOT}$ is invariant under $\symsub{m}{z}$ since neither $m_z$ nor $\vec e_z$ appear in \autoref{eq:heff}. 

\prlsection{Micromagnetics}
The magnetization dynamics can be described by the semiclassical Landau--Lifshitz--Gilbert (LLG) equation
\begin{equation}
\vec{\dot{m}} = -\abs{\gamma}\vec{m}\times\vec{H}_\text{eff} + \alpha\vec{m}\times\vec{\dot{m}} + \vec{\tau} \, .
\label{eq:LLG}
\end{equation}
Here, $\alpha>0$ is the dimensionless Gilbert damping parameter, $\vec{H}_\text{eff} = -M_s^{-1} \delta f\left[\vec{m}\right]/\delta\vec{m}$ is an effective magnetic field that describes the magnetization direction $\vec{m}$ that minimizes the free energy density functional $f\left[\vec{m}\right]$, and $M_s$ is the saturation magnetization.
Interestingly, we note that a functional exists that generates the effective SOT field in \autoref{eq:heff}, which is given by
\begin{equation}
f_\text{SOT}\left[\vec{m}\right] = M_s\Gamma_0\left[J_y m_x m_y-\frac{1}{2}J_x\left(m_x^2-m_y^2\right)\right].
\label{eq:f_SOT}
\end{equation}
The out-of-equilibrium current-induced SOT can therefore be absorbed into an effective field $\vec{\tilde{H}}_\text{eff}$ that minimizes the \textit{effective} free energy density $f_\text{eff}\left[\vec{m}\right]=f\left[\vec{m}\right]+f_\text{SOT}\left[\vec{m}\right]$.

The 2D ferromagnet \ce{Fe3GeTe2} is a uniaxial ferromagnet with an out-of-plane easy axis~\cite{Deng2018a,Fei2018,Zhuang:prb:2016}.
The contribution of the dipole--dipole interaction to the spin-wave spectrum can be neglected for a monolayer system~\cite{Chartoryzhskii1976,Kalinikos1980,Kalinikos1981a,Kalinikos1986,Kalinikos1990}.
If we consider a spatially uniform magnetization and use a spherical basis, ${\left(m_x,m_y,m_z\right)} = {\left(\sin\theta\cos\phi,\sin\theta\sin\phi,\cos\theta\right)}$, the effective free energy becomes
\begin{equation}
f_\text{eff}\left[\theta,\phi\right] = -\frac{M_s}{2}\left[K_z\cos^2\theta+\Gamma_0\abs{J}\sin^2\theta\cos\left(2\phi+\phi_J\right)\right].
\label{eq:f_eff_spherical}
\end{equation}
Here, $K_z>0$ is the out-of-plane anisotropy constant, and $\abs{J}$ and $\phi_J=\arctan\left(J_x/J_y\right)$ are the magnitude and azimuthal angle of the applied current, respectively.
From this, we find that the SOT effectively acts as in-plane magnetocrystalline anisotropies. 
The anisotropy originating from the SOT always comes in a pair of perpendicular easy and hard axes. 
The directions of the anisotropy axes depend on the direction of the applied current.
For weak currents ($\abs{\Gamma_0 J}<K_z$), the magnetization of \ce{Fe3GeTe2} remains out of plane ($\theta=0,\pi$).
However, for sufficiently strong currents ($\abs{\Gamma_0 J}>K_z$), an in-plane configuration of the magnetization becomes more energetically favorable. 
Assuming that $\Gamma_0>0$, the effective free energy is then minimized by $\theta=\pi/2$ and $\phi=n\pi-\phi_J/2$ ($n=0, 1, 2, \ldots$).
When $\Gamma_0<0$, the easy and hard axes are interchanged, and the minima are $\phi=\left(n+1/2\right)\pi-\phi_J/2$.
The easy and hard axes also interchange upon reversal of the applied current.

\prlsection{Magnon gap}
Because the SOT can effectively be considered  a current-controlled magnetocrystalline anisotropy, we can electrically control the magnon gap in \ce{Fe3GeTe2}. 
The magnon gap is governed by the energy difference between the out-of-plane and in-plane magnetization configurations, \ie\, $\abs{K_z-\abs{\Gamma_0 J}}$.
At the critical current $\abs{J_c}=K_z/\abs{\Gamma_0}$, the magnon gap vanishes as the magnetic easy axis transitions from an out-of-plane axis to an in-plane axis.
Exactly at this transition point, we obtain a magnetic easy plane.
Below the critical current, the magnon gap decreases monotonically with the applied current, whereas it increases monotonically above the critical current.
The ability to electrically tune the magnon gap in a 2D magnetic material opens the door for exploring a wide variety of effects in magnetism in two dimensions.

\prlsection{Curie temperature}
The first effect that is characteristic of a two-dimensional system that we will now illustrate is the dependence of the Curie temperature on the magnon gap. Because the Curie temperature in 2D is primarily governed by the magnon gap, unlike in 3D~\cite{Auerbach1994}, we will study its behavior as we tune the SOT-controlled magnon gap through the transition from an out-of-plane easy axis to an in-plane easy axis.
To illustrate the basic aspects of current control of the Curie temperature, we make a few simplifications to reduce the number of free parameters and the complexity of the calculations. \ce{Fe3GeTe2} is an itinerant ferromagnet, and its magnetic interactions are therefore described by the Stoner model~\cite{Zhuang:prb:2016}. The Stoner model can in our system be transformed into an RKKY exchange interaction between the iron atoms~\cite{Prange:prb:1979}.
We assume that the exchange interaction in an \ce{Fe3GeTe2} monolayer has a finite range and therefore obeys the Mermin--Wagner theorem.
To simplify the calculations, we replace the Stoner/RKKY exchange interaction by a simple nearest-neighbor interaction between the \ce{Fe^{II}} and  \ce{Fe^{III}} atoms (\ie\, there is no exchange interaction within each sublattice or between the two different \ce{Fe^{III}} sublattices).
This will also obey the Mermin--Wagner theorem, and this system will consequently also exhibit the same qualitative dependence on the magnon gap as other finite-range interactions.
We also assume that the magnetic anisotropy constants are identical at all sites. Consequently, we consider the model Hamiltonian
\begin{align}
\nonumber \mathcal{H} = &-\frac{\varepsilon_J}{2\hbar^2}\sum_{\vec r}\sum_{\vec\delta} \vec{S}_{\vec{r}}\cdot\vec{S}_{\vec{r}+\vec{\delta}} - \frac{\varepsilon_z}{2\hbar^2}\sum_{\vec{r}}\left(S_{\vec{r},z}\right)^2 \\
&- \frac{\varepsilon_x}{2\hbar^2}\sum_{\vec{r}}\left[\left(S_{\vec{r},x}\right)^2-\left(S_{\vec{r},y}\right)^2\right] \, .
\label{eq:Hamiltonian}
\end{align}
Here, $\varepsilon_J>0$ is an energy constant that describes the nearest-neighbor exchange interactions of spins separated by $\vec{\delta}$, $\varepsilon_z>0$ is an energy constant that describes the out-of-plane anisotropy, and $\varepsilon_x\propto \Gamma_0 J_x>0$ is an energy constant that describes the effective in-plane anisotropies caused by the SOT.
$S_{\vec{r},i}$ ($i=x,y,z$) describes the $i$-th component of the spin operator located at position $\vec{r}$.
We split the \ce{Fe3GeTe2} monolayer into three distinct sublattices: one for the \ce{Fe^{II}} atoms, one for the \ce{Fe^{III}} atoms at $z=+b$, and one for the \ce{Fe^{III}} atoms at $z=-b$.

We proceed by performing a Holstein--Primakoff transformation of the spin operators around the equilibrium spin direction.
This is in the $z$ direction below the critical current $J_c$ and along the $x$ direction above the critical current.
Because of the anomalous Hall effect in \ce{Fe3GeTe2}~\cite{Deng2018a,Tan2018,Nagaosa2010}, applying the current exactly along the $x$ direction can be experimentally challenging.
However, as can be deduced from \autoref{eq:f_eff_spherical}, a scenario in which the current is applied in a different direction can be achieved by a rotation of the unit cell or Brillouin zone.
Since it is the magnons closest to the $\Gamma$ point that dominate the calculation of the Curie temperature, we expect the results to be very similar for an off-axis current.

In our calculations, we keep terms to the second order in the Holstein--Primakoff magnon operators. 
We expect this to be a good qualitative approximation, although it will not be a very good quantitative approximation because the magnon population diverges at the critical point.
However, keeping terms to, for instance, the fourth order in the magnon operators to include magnon--magnon interactions~\cite{Gong2017} would be complicated because \autoref{eq:Hamiltonian} does not conserve the magnon number for finite currents.

Following the Holstein--Primakoff transformation, we perform a Fourier transformation of the magnon operators to momentum space.
We then diagonalize the Hamiltonian by a Bogoliubov transformation such that it takes the form~\cite{Note1}
\begin{equation}
\mathcal{H} = \sum_{\vec{k},\mu}\varepsilon_{\vec{k},\mu}\alpha_{\vec{k},\mu}^\dagger \alpha_{\vec{k},\mu} \, .
\label{eq:Diagonal_Hamiltonian}
\end{equation}
Here, the operator $\alpha_{\vec{k},\mu}^{(\dagger)}$ annihilates (creates) an eigenmagnon with a momentum $\vec{k}$ and energy $\varepsilon_{\vec{k},\mu}$.
There are three different modes ($\mu = \text{I},\, \text{II},\, \text{III}$) of the eigenmagnons.
We have imposed the constraint on the Bogoliubov transformation that the new operators have to satisfy bosonic commutation relations: ${[\alpha_{\vec{k},\mu}, \alpha_{\vec{k'},\mu'}^\dagger ]} = \delta_{\vec{k}\vec{k'}}\delta_{\mu\mu'}$.

From the energy spectrum of the eigenmagnons in \ce{Fe3GeTe2}, we can estimate the Curie temperature $T_c$.
To determine $T_c$, we use the fact that the magnetization along the equilibrium direction of the spins vanishes at this temperature. 
Because we consider a monolayer system, we only have magnons with in-plane momenta.
Balancing the magnetic moments, we find the constraint
\begin{equation}
\sum_\nu s_\nu - \sum_\mu\frac{1}{A_{\text{BZ}}}\int_{A_{\text{BZ}}}\drm^2k  \frac{S_{\vec{k},\mu}/\hbar}{\exp\left(\varepsilon_{\vec{k},\mu}/k_\text{B}T_c\right)-1} = 0 \, .
\label{eq:Tc_Criterion}
\end{equation}
Here, $s_\nu$ is the dimensionless spin number of the magnetic moments in sublattice $\nu$ (where $\nu=2$ for the \ce{Fe^{II}} atoms, and $\nu=3_\pm$ for the \ce{Fe^{III}} atoms located at $z=\pm b$), and $A_\text{BZ}= \sqrt{3}\pi^2/(2a^2)$ is the (reciprocal) area of the first Brillouin zone.
$S_{\vec{k},\mu}$ is the spin of the eigenmagnons, which is \emph{not} an integer for finite SOT because of magnon squeezing~\cite{Kamra:prb:2017}.
The spin of the eigenmagnons depends on the parameters of the Bogoliubov transformation and is given in the Supplementary Material~\cite{Note1}.

\begin{figure}[t]
  \includegraphics[width=.95\columnwidth]{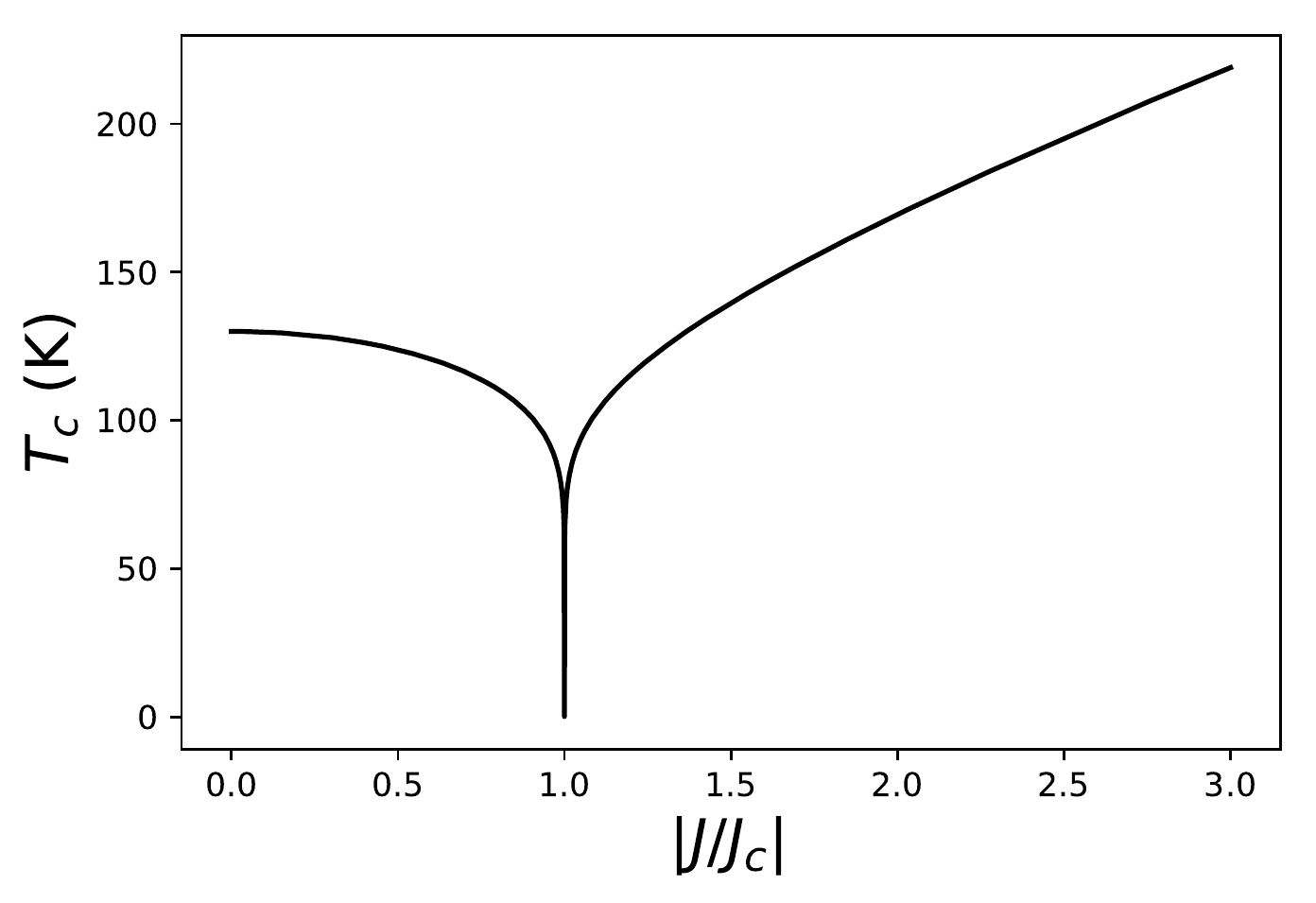}
  \caption{\label{fig:Tc}
  Numerical calculation of $T_c$ for a spontaneous magnetization based on a simple linear response model of the magnon spectrum. 
  The result is identical for any direction of the applied current $\vec J$.
  Below $\abs{J_c}$, the magnetization is along the $z$ axis, whereas above $\abs{J_c}$, the magnetization is along an in-plane axis determined by the direction of the applied current.
  }
\end{figure}

We can now calculate the Curie temperature numerically based on \autoref{eq:Tc_Criterion}.
In our calculations, we set the out-of-plane anisotropy constant to be $\varepsilon_z = 0.335$~meV~\cite{Zhuang:prb:2016}.
The value of the nearest-neighbor exchange coupling is set to be $\varepsilon_J = 0.705$~meV to reproduce the experimental $T_c$ of a monolayer of $\sim 130$~K~\cite{Fei2018} (note, however, that a different experiment determined the $T_c$ of a monolayer to be $\sim 68$~K~\cite{Deng2018a}).
The real value of $\varepsilon_J$ is in all likelihood larger~\cite{Deng2018a} because the linear response method typically overestimates $T_c$.
The dimensionless spin numbers $s_\nu$ for the spins in sublattice $\nu$ are $s_2=2$ and $s_{3_-},\, s_{3_+}=5/2$~\cite{Rodriguez:JPC:1996}.
We plot the Curie temperature as a function of the applied current in \autoref{fig:Tc}.

Because we only kept terms to the second order in the magnon operators, we do not expect that our calculation of $T_c$ will be quantitatively correct.
However, the qualitative features of our result appear to be physically reasonable.
When we apply a SOT below the critical current $\abs{J_c}$, we effectively reduce the magnon gap by creating a pair of easy and hard axes perpendicular to the out-of-plane magnetization.
Because the Curie temperature in 2D materials is governed by the magnon gap, this also reduces $T_c$.
At the critical current strength, we obtain a continuous symmetry in the form of an easy plane when the in-plane easy axis induced by the SOT becomes equal to the out-of-plane magnetocrystalline anisotropy.
Because of the Mermin--Wagner theorem, there can be no long-range magnetic order at finite temperatures in this scenario, and $T_c$ drops to zero.
Above the critical current, we now increase the magnon gap for an in-plane magnetization configuration, and $T_c$ increases accordingly.
$T_c$ will then saturate at the Curie temperature of the Ising model for large currents,  which our model does not capture~\cite{Torelli}.

In addition to the current affecting the Curie temperature through a SOT, the current will also increase the temperature in the material due to Joule heating, which needs to be taken into account when measuring the Curie temperature of the material.
The Joule heating increases quadratically with the applied current. 
Conversely, the SOT is linear in the applied current, but its effect on the Curie temperature depends on whether we are above or below the critical current. 
Consequently, if the critical current is sufficiently small, then the effect of the SOT will dominate that of the Joule heating.
In this case, the magnetic ordering exhibits reentrant behavior as a function of the applied current.
Notably, above the critical current, when the magnetization is in the plane, the easy and hard axes are interchanged upon reversal of the current direction.
A reversal of the applied current would therefore lead to a $90^\circ$ rotation of the magnetization.

\prlsection{2D $XY$ model}
 Although the spontaneous magnetization vanishes for finite temperatures at the critical current density $\abs{J_c}$, this regime remains an interesting region for studying the magnetic properties.
At the critical current density ($\abs{\varepsilon_x}=\varepsilon_z$), the model in \autoref{eq:Hamiltonian} becomes, quite remarkably, a 2D easy-plane ferromagnet, where the easy plane is perpendicular to the plane of the monolayer. 
Therefore, at this current density, the model features a critical phenomenon in the universality class of the 2D $XY$ model. Consequently, the system has a topological phase transition rather than the more conventional phase transition of the 2D Ising model~\cite{Onsager}. The 2D Ising universality class  falls within the framework of the Landau--Ginzburg--Wilson paradigm of phase transitions of an order--disorder transition monitored by a local order parameter~\cite{Landau-Ginzburg,Wilson-Kogut}. The spin--spin correlation length diverges from above and below $T_c$ as $\xi \sim |T-T_c|^{-\nu}$, where $\nu$ is a universal critical exponent. There is true long-range order in the low-temperature phase, short-range order in the high-temperature phase, and power-law spin--spin correlations precisely at the critical point. In contrast, the 2D $XY$ model features a genuine phase transition with \textit{no local order parameter}. 
At this phase transition, the spin--spin correlation length diverges as $\xi \sim \exp(\mathrm{const}/\sqrt{T-T_{\mathrm{BKT}}})$ from the high-temperature side only~\cite{KosterlitzThouless}, where $T_{\mathrm{BKT}}$ is the critical temperature of the BKT transition.
The high-temperature phase has short-range order, and \textit{the entire low-temperature phase} is critical with a spin--spin correlation function featuring a nonuniversal temperature-dependent anomalous dimension $\eta$, $\langle \vec{S}_{\vec{r}}\cdot\vec{S}_{\vec{r}^{\prime}}\rangle \sim 1/|{\bf r}-{\bf r}^{\prime}|^{\eta}$ \cite{KosterlitzThouless}.  

In 2D \ce{Fe3GeTe2}, we may realize this type of highly nontrivial behavior by tuning the electric current to the critical value and then drive the system through the phase transition by varying the temperature. 
Moreover, below the BKT transition, the temperature dependence of the nonuniversal anomalous dimension $\eta$ of the 2D $XY$ model can be mapped by varying the temperature and measuring the spin--spin correlation function by polarized small-angle neutron scattering, which is particularly well suited for ultrathin films~\cite{Maurer}. The present system is also amenable to studying the universal anomalous dimension of the 2D Ising-model at $T=T_c$, ${\eta=1/4}$~\cite{Itzykson-Drouffe}. The prediction for the 2D $XY$ model,
$\eta =  k_B T/4 \pi J$~\cite{KosterlitzThouless}, where $J$ is the effective exchange coupling and $k_B$ is Boltzmann's constant,  has  not been tested in real 2D magnetic systems to our knowledge.   

Examples of real physical systems with this level of control over  such phenomena are very rare, particularly for systems where the phenomena are accessible at relatively elevated temperatures. The most well-known example is superfluidity in thin films of $^4\mathrm{He}$, where the BKT transition occurs below $\SI{1.2}{K}$~\cite{BishopReppy}. In that context, the remarkable prediction and experimental verification of a universal jump in the superfluid density of the system~\cite{nelson1977universal,BishopReppy} is also worth noting. We expect the corresponding physics of a universal jump in the spin stiffness of the system to occur at liquid nitrogen or oxygen  temperatures in the system studied here.
The spin stiffness may be measured in spin-wave resonance experiments~\cite{Golosovsky}. 
Furthermore, and in contrast to our present case, $\eta$ is not experimentally accessible in superfluid thin films of $^4\mathrm{He}$. 

The parameter $\Gamma_0$ determines the magnitude of the critical current and thus the accessibility of the effects that we discuss. This value cannot be obtained purely from symmetry considerations but rather needs to be determined experimentally or by \textit{ab initio} calculations.
In light of the exciting physics that can be realized and the flexibility of the system, determining its value would be very interesting. 
Based on the strong magnetic anisotropy of the material, we believe that the spin--orbit coupling is sufficiently strong. 
Paired with the observation that SOTs are typically sufficiently large to induce magnetization switching in other materials~\cite{Manchon}, we have reason to believe that reentrant magnetism and topological phase transitions can be experimentally observed in \ce{Fe3GeTe2}. 

\begin{acknowledgments}
The authors thank Alireza Qaiumzadeh for helpful discussions.
We gratefully acknowledge funding via the ``Outstanding Academic Fellows'' program at NTNU, the Research Council of Norway Grant No.\ 240806, 239926, and 250985, and the Research Council of Norway through its Centres of Excellence funding scheme, Project No.\ 262633, ``QuSpin''.

Ø.\,J.\ and V.\,R.\ contributed equally to this work.
\end{acknowledgments}


\begin{thebibliography}{48}%
\makeatletter
\providecommand \@ifxundefined [1]{%
 \@ifx{#1\undefined}
}%
\providecommand \@ifnum [1]{%
 \ifnum #1\expandafter \@firstoftwo
 \else \expandafter \@secondoftwo
 \fi
}%
\providecommand \@ifx [1]{%
 \ifx #1\expandafter \@firstoftwo
 \else \expandafter \@secondoftwo
 \fi
}%
\providecommand \natexlab [1]{#1}%
\providecommand \enquote  [1]{``#1''}%
\providecommand \bibnamefont  [1]{#1}%
\providecommand \bibfnamefont [1]{#1}%
\providecommand \citenamefont [1]{#1}%
\providecommand \href@noop [0]{\@secondoftwo}%
\providecommand \href [0]{\begingroup \@sanitize@url \@href}%
\providecommand \@href[1]{\@@startlink{#1}\@@href}%
\providecommand \@@href[1]{\endgroup#1\@@endlink}%
\providecommand \@sanitize@url [0]{\catcode `\\12\catcode `\$12\catcode
  `\&12\catcode `\#12\catcode `\^12\catcode `\_12\catcode `\%12\relax}%
\providecommand \@@startlink[1]{}%
\providecommand \@@endlink[0]{}%
\providecommand \url  [0]{\begingroup\@sanitize@url \@url }%
\providecommand \@url [1]{\endgroup\@href {#1}{\urlprefix }}%
\providecommand \urlprefix  [0]{URL }%
\providecommand \Eprint [0]{\href }%
\providecommand \doibase [0]{http://dx.doi.org/}%
\providecommand \selectlanguage [0]{\@gobble}%
\providecommand \bibinfo  [0]{\@secondoftwo}%
\providecommand \bibfield  [0]{\@secondoftwo}%
\providecommand \translation [1]{[#1]}%
\providecommand \BibitemOpen [0]{}%
\providecommand \bibitemStop [0]{}%
\providecommand \bibitemNoStop [0]{.\EOS\space}%
\providecommand \EOS [0]{\spacefactor3000\relax}%
\providecommand \BibitemShut  [1]{\csname bibitem#1\endcsname}%
\let\auto@bib@innerbib\@empty
\bibitem [{\citenamefont {Onsager}(1944)}]{Onsager}%
  \BibitemOpen
  \bibfield  {author} {\bibinfo {author} {\bibfnamefont {L.}~\bibnamefont
  {Onsager}},\ }\bibfield  {title} {\enquote {\bibinfo {title} {Crystal
  statistics. {I. A} two-dimensional model with an order-disorder
  transition},}\ }\href {\doibase 10.1103/PhysRev.65.117} {\bibfield  {journal}
  {\bibinfo  {journal} {Phys. Rev.}\ }\textbf {\bibinfo {volume} {65}},\
  \bibinfo {pages} {117--149} (\bibinfo {year} {1944})}\BibitemShut {NoStop}%
\bibitem [{\citenamefont {{Berezinski{\v i}}}(1971)}]{Berezinskii1}%
  \BibitemOpen
  \bibfield  {author} {\bibinfo {author} {\bibfnamefont {V.~L.}\ \bibnamefont
  {{Berezinski{\v i}}}},\ }\bibfield  {title} {\enquote {\bibinfo {title}
  {Destruction of long-range order in one-dimensional and two-dimensional
  systems having a continuous symmetry group {I. C}lassical systems},}\
  }\href@noop {} {\bibfield  {journal} {\bibinfo  {journal} {Soviet Phys.
  JETP}\ }\textbf {\bibinfo {volume} {32}},\ \bibinfo {pages} {493} (\bibinfo
  {year} {1971})}\BibitemShut {NoStop}%
\bibitem [{\citenamefont {{Berezinski{\v i}}}(1972)}]{Berezinskii2}%
  \BibitemOpen
  \bibfield  {author} {\bibinfo {author} {\bibfnamefont {V.~L.}\ \bibnamefont
  {{Berezinski{\v i}}}},\ }\bibfield  {title} {\enquote {\bibinfo {title}
  {Destruction of long-range order in one-dimensional and two-dimensional
  systems possessing a continuous symmetry group. {II. Q}uantum systems},}\
  }\href@noop {} {\bibfield  {journal} {\bibinfo  {journal} {Soviet Phys.
  JETP}\ }\textbf {\bibinfo {volume} {34}},\ \bibinfo {pages} {610} (\bibinfo
  {year} {1972})}\BibitemShut {NoStop}%
\bibitem [{\citenamefont {Kosterlitz}\ and\ \citenamefont
  {Thouless}(1973)}]{KosterlitzThouless}%
  \BibitemOpen
  \bibfield  {author} {\bibinfo {author} {\bibfnamefont {J.~M.}\ \bibnamefont
  {Kosterlitz}}\ and\ \bibinfo {author} {\bibfnamefont {D.~J.}\ \bibnamefont
  {Thouless}},\ }\bibfield  {title} {\enquote {\bibinfo {title} {Ordering,
  metastability and phase transitions in two-dimensional systems},}\ }\href
  {http://stacks.iop.org/0022-3719/6/i=7/a=010} {\bibfield  {journal} {\bibinfo
   {journal} {J. Phys. C: Solid State}\ }\textbf {\bibinfo {volume} {6}},\
  \bibinfo {pages} {1181} (\bibinfo {year} {1973})}\BibitemShut {NoStop}%
\bibitem [{\citenamefont {Novoselov}\ \emph {et~al.}(2004)\citenamefont
  {Novoselov}, \citenamefont {Geim}, \citenamefont {Morozov}, \citenamefont
  {Jiang}, \citenamefont {Zhang}, \citenamefont {Dubonos}, \citenamefont
  {Grigorieva},\ and\ \citenamefont {Firsov}}]{Novoselov}%
  \BibitemOpen
  \bibfield  {author} {\bibinfo {author} {\bibfnamefont {K.~S.}\ \bibnamefont
  {Novoselov}}, \bibinfo {author} {\bibfnamefont {A.~K.}\ \bibnamefont {Geim}},
  \bibinfo {author} {\bibfnamefont {S.~V.}\ \bibnamefont {Morozov}}, \bibinfo
  {author} {\bibfnamefont {D.}~\bibnamefont {Jiang}}, \bibinfo {author}
  {\bibfnamefont {Y.}~\bibnamefont {Zhang}}, \bibinfo {author} {\bibfnamefont
  {S.~V.}\ \bibnamefont {Dubonos}}, \bibinfo {author} {\bibfnamefont {I.~V.}\
  \bibnamefont {Grigorieva}}, \ and\ \bibinfo {author} {\bibfnamefont {A.~A.}\
  \bibnamefont {Firsov}},\ }\bibfield  {title} {\enquote {\bibinfo {title}
  {Electric field effect in atomically thin carbon films},}\ }\href {\doibase
  10.1126/science.1102896} {\bibfield  {journal} {\bibinfo  {journal}
  {Science}\ }\textbf {\bibinfo {volume} {306}},\ \bibinfo {pages} {666--669}
  (\bibinfo {year} {2004})}\BibitemShut {NoStop}%
\bibitem [{\citenamefont {Mermin}\ and\ \citenamefont
  {Wagner}(1966)}]{MerminWagner}%
  \BibitemOpen
  \bibfield  {author} {\bibinfo {author} {\bibfnamefont {N.~D.}\ \bibnamefont
  {Mermin}}\ and\ \bibinfo {author} {\bibfnamefont {H.}~\bibnamefont
  {Wagner}},\ }\bibfield  {title} {\enquote {\bibinfo {title} {Absence of
  ferromagnetism or antiferromagnetism in one- or two-dimensional isotropic
  {Heisenberg} models},}\ }\href {\doibase 10.1103/PhysRevLett.17.1133}
  {\bibfield  {journal} {\bibinfo  {journal} {Phys. Rev. Lett.}\ }\textbf
  {\bibinfo {volume} {17}},\ \bibinfo {pages} {1133--1136} (\bibinfo {year}
  {1966})}\BibitemShut {NoStop}%
\bibitem [{\citenamefont {Burch}\ \emph {et~al.}(2018)\citenamefont {Burch},
  \citenamefont {Mandrus},\ and\ \citenamefont {Park}}]{Burch2018}%
  \BibitemOpen
  \bibfield  {author} {\bibinfo {author} {\bibfnamefont {K.~S.}\ \bibnamefont
  {Burch}}, \bibinfo {author} {\bibfnamefont {D.}~\bibnamefont {Mandrus}}, \
  and\ \bibinfo {author} {\bibfnamefont {J.-G.}\ \bibnamefont {Park}},\
  }\bibfield  {title} {\enquote {\bibinfo {title} {{Magnetism in
  two-dimensional van der Waals materials}},}\ }\href {\doibase
  10.1038/s41586-018-0631-z} {\bibfield  {journal} {\bibinfo  {journal}
  {Nature}\ }\textbf {\bibinfo {volume} {563}},\ \bibinfo {pages} {47--52}
  (\bibinfo {year} {2018})}\BibitemShut {NoStop}%
\bibitem [{\citenamefont {Lee}\ \emph {et~al.}(2016)\citenamefont {Lee},
  \citenamefont {Lee}, \citenamefont {Ryoo}, \citenamefont {Kang},
  \citenamefont {Kim}, \citenamefont {Kim}, \citenamefont {Park}, \citenamefont
  {Park},\ and\ \citenamefont {Cheong}}]{Lee2016}%
  \BibitemOpen
  \bibfield  {author} {\bibinfo {author} {\bibfnamefont {J.~U.}\ \bibnamefont
  {Lee}}, \bibinfo {author} {\bibfnamefont {S.}~\bibnamefont {Lee}}, \bibinfo
  {author} {\bibfnamefont {J.~H.}\ \bibnamefont {Ryoo}}, \bibinfo {author}
  {\bibfnamefont {S.}~\bibnamefont {Kang}}, \bibinfo {author} {\bibfnamefont
  {T.~Y.}\ \bibnamefont {Kim}}, \bibinfo {author} {\bibfnamefont
  {P.}~\bibnamefont {Kim}}, \bibinfo {author} {\bibfnamefont {C.-H.}\
  \bibnamefont {Park}}, \bibinfo {author} {\bibfnamefont {J.-G.}\ \bibnamefont
  {Park}}, \ and\ \bibinfo {author} {\bibfnamefont {H.}~\bibnamefont
  {Cheong}},\ }\bibfield  {title} {\enquote {\bibinfo {title} {{Ising-type
  magnetic ordering in atomically thin \ce{FePS3}}},}\ }\href {\doibase
  10.1021/acs.nanolett.6b03052} {\bibfield  {journal} {\bibinfo  {journal}
  {Nano Lett.}\ }\textbf {\bibinfo {volume} {16}},\ \bibinfo {pages}
  {7433--7438} (\bibinfo {year} {2016})}\BibitemShut {NoStop}%
\bibitem [{\citenamefont {Gong}\ \emph {et~al.}(2017)\citenamefont {Gong},
  \citenamefont {Li}, \citenamefont {Li}, \citenamefont {Ji}, \citenamefont
  {Stern}, \citenamefont {Xia}, \citenamefont {Cao}, \citenamefont {Bao},
  \citenamefont {Wang}, \citenamefont {Wang}, \citenamefont {Qiu},
  \citenamefont {Cava}, \citenamefont {Louie}, \citenamefont {Xia},\ and\
  \citenamefont {Zhang}}]{Gong2017}%
  \BibitemOpen
  \bibfield  {author} {\bibinfo {author} {\bibfnamefont {C.}~\bibnamefont
  {Gong}}, \bibinfo {author} {\bibfnamefont {L.}~\bibnamefont {Li}}, \bibinfo
  {author} {\bibfnamefont {Z.}~\bibnamefont {Li}}, \bibinfo {author}
  {\bibfnamefont {H.}~\bibnamefont {Ji}}, \bibinfo {author} {\bibfnamefont
  {A.}~\bibnamefont {Stern}}, \bibinfo {author} {\bibfnamefont
  {Y.}~\bibnamefont {Xia}}, \bibinfo {author} {\bibfnamefont {T.}~\bibnamefont
  {Cao}}, \bibinfo {author} {\bibfnamefont {W.}~\bibnamefont {Bao}}, \bibinfo
  {author} {\bibfnamefont {C.}~\bibnamefont {Wang}}, \bibinfo {author}
  {\bibfnamefont {Y.}~\bibnamefont {Wang}}, \bibinfo {author} {\bibfnamefont
  {Z.~Q.}\ \bibnamefont {Qiu}}, \bibinfo {author} {\bibfnamefont {R.~J.}\
  \bibnamefont {Cava}}, \bibinfo {author} {\bibfnamefont {S.~G.}\ \bibnamefont
  {Louie}}, \bibinfo {author} {\bibfnamefont {J.}~\bibnamefont {Xia}}, \ and\
  \bibinfo {author} {\bibfnamefont {X.}~\bibnamefont {Zhang}},\ }\bibfield
  {title} {\enquote {\bibinfo {title} {{Discovery of intrinsic ferromagnetism
  in two-dimensional van der Waals crystals}},}\ }\href {\doibase
  10.1038/nature22060} {\bibfield  {journal} {\bibinfo  {journal} {Nature}\
  }\textbf {\bibinfo {volume} {546}},\ \bibinfo {pages} {265--269} (\bibinfo
  {year} {2017})}\BibitemShut {NoStop}%
\bibitem [{\citenamefont {Huang}\ \emph {et~al.}(2017)\citenamefont {Huang},
  \citenamefont {Clark}, \citenamefont {Navarro-Moratalla}, \citenamefont
  {Klein}, \citenamefont {Cheng}, \citenamefont {Seyler}, \citenamefont
  {Zhong}, \citenamefont {Schmidgall}, \citenamefont {McGuire}, \citenamefont
  {Cobden}, \citenamefont {Yao}, \citenamefont {Xiao}, \citenamefont
  {Jarillo-Herrero},\ and\ \citenamefont {Xu}}]{Huang2017}%
  \BibitemOpen
  \bibfield  {author} {\bibinfo {author} {\bibfnamefont {B.}~\bibnamefont
  {Huang}}, \bibinfo {author} {\bibfnamefont {G.}~\bibnamefont {Clark}},
  \bibinfo {author} {\bibfnamefont {E.}~\bibnamefont {Navarro-Moratalla}},
  \bibinfo {author} {\bibfnamefont {D.~R.}\ \bibnamefont {Klein}}, \bibinfo
  {author} {\bibfnamefont {R.}~\bibnamefont {Cheng}}, \bibinfo {author}
  {\bibfnamefont {K.~L.}\ \bibnamefont {Seyler}}, \bibinfo {author}
  {\bibfnamefont {D.}~\bibnamefont {Zhong}}, \bibinfo {author} {\bibfnamefont
  {E.}~\bibnamefont {Schmidgall}}, \bibinfo {author} {\bibfnamefont {M.~A.}\
  \bibnamefont {McGuire}}, \bibinfo {author} {\bibfnamefont {D.~H.}\
  \bibnamefont {Cobden}}, \bibinfo {author} {\bibfnamefont {W.}~\bibnamefont
  {Yao}}, \bibinfo {author} {\bibfnamefont {D.}~\bibnamefont {Xiao}}, \bibinfo
  {author} {\bibfnamefont {P.}~\bibnamefont {Jarillo-Herrero}}, \ and\ \bibinfo
  {author} {\bibfnamefont {X.}~\bibnamefont {Xu}},\ }\bibfield  {title}
  {\enquote {\bibinfo {title} {{Layer-dependent ferromagnetism in a van der
  Waals crystal down to the monolayer limit}},}\ }\href {\doibase
  10.1038/nature22391} {\bibfield  {journal} {\bibinfo  {journal} {Nature}\
  }\textbf {\bibinfo {volume} {546}},\ \bibinfo {pages} {270--273} (\bibinfo
  {year} {2017})}\BibitemShut {NoStop}%
\bibitem [{\citenamefont {Bonilla}\ \emph {et~al.}(2018)\citenamefont
  {Bonilla}, \citenamefont {Kolekar}, \citenamefont {Ma}, \citenamefont {Diaz},
  \citenamefont {Kalappattil}, \citenamefont {Das}, \citenamefont {Eggers},
  \citenamefont {Gutierrez}, \citenamefont {Phan},\ and\ \citenamefont
  {Batzill}}]{Bonilla2018}%
  \BibitemOpen
  \bibfield  {author} {\bibinfo {author} {\bibfnamefont {M.}~\bibnamefont
  {Bonilla}}, \bibinfo {author} {\bibfnamefont {S.}~\bibnamefont {Kolekar}},
  \bibinfo {author} {\bibfnamefont {Y.}~\bibnamefont {Ma}}, \bibinfo {author}
  {\bibfnamefont {H.~C.}\ \bibnamefont {Diaz}}, \bibinfo {author}
  {\bibfnamefont {V.}~\bibnamefont {Kalappattil}}, \bibinfo {author}
  {\bibfnamefont {R.}~\bibnamefont {Das}}, \bibinfo {author} {\bibfnamefont
  {T.}~\bibnamefont {Eggers}}, \bibinfo {author} {\bibfnamefont {H.~R.}\
  \bibnamefont {Gutierrez}}, \bibinfo {author} {\bibfnamefont {M.-H.}\
  \bibnamefont {Phan}}, \ and\ \bibinfo {author} {\bibfnamefont
  {M.}~\bibnamefont {Batzill}},\ }\bibfield  {title} {\enquote {\bibinfo
  {title} {{Strong room-temperature ferromagnetism in \ce{VSe2} monolayers on
  van der Waals substrates}},}\ }\href {\doibase 10.1038/s41565-018-0063-9}
  {\bibfield  {journal} {\bibinfo  {journal} {Nat. Nanotechnol.}\ }\textbf
  {\bibinfo {volume} {13}},\ \bibinfo {pages} {289--293} (\bibinfo {year}
  {2018})}\BibitemShut {NoStop}%
\bibitem [{\citenamefont {O'Hara}\ \emph {et~al.}(2018)\citenamefont {O'Hara},
  \citenamefont {Zhu}, \citenamefont {Trout}, \citenamefont {Ahmed},
  \citenamefont {Yunqiu}, \citenamefont {Lee}, \citenamefont {Brenner},
  \citenamefont {Rajan}, \citenamefont {Gupta}, \citenamefont {McComb},\ and\
  \citenamefont {Kawakami}}]{OHara}%
  \BibitemOpen
  \bibfield  {author} {\bibinfo {author} {\bibfnamefont {D.~J.}\ \bibnamefont
  {O'Hara}}, \bibinfo {author} {\bibfnamefont {T.}~\bibnamefont {Zhu}},
  \bibinfo {author} {\bibfnamefont {A.~H.}\ \bibnamefont {Trout}}, \bibinfo
  {author} {\bibfnamefont {A.~S.}\ \bibnamefont {Ahmed}}, \bibinfo {author}
  {\bibfnamefont {L.}~\bibnamefont {Yunqiu}}, \bibinfo {author} {\bibfnamefont
  {C.~H.}\ \bibnamefont {Lee}}, \bibinfo {author} {\bibfnamefont {M.~R.}\
  \bibnamefont {Brenner}}, \bibinfo {author} {\bibfnamefont {S.}~\bibnamefont
  {Rajan}}, \bibinfo {author} {\bibfnamefont {J.~A.}\ \bibnamefont {Gupta}},
  \bibinfo {author} {\bibfnamefont {D.~W.}\ \bibnamefont {McComb}}, \ and\
  \bibinfo {author} {\bibfnamefont {R.~K.}\ \bibnamefont {Kawakami}},\
  }\bibfield  {title} {\enquote {\bibinfo {title} {{Room temperature intrinsic
  ferromagnetism in epitaxial manganese selenide films in the monolayer
  limit}},}\ }\href {\doibase 10.1021/acs.nanolett.8b00683} {\bibfield
  {journal} {\bibinfo  {journal} {Nano Lett.}\ }\textbf {\bibinfo {volume}
  {18}},\ \bibinfo {pages} {3125--3131} (\bibinfo {year} {2018})}\BibitemShut
  {NoStop}%
\bibitem [{\citenamefont {Fei}\ \emph {et~al.}(2018)\citenamefont {Fei},
  \citenamefont {Huang}, \citenamefont {Malinowski}, \citenamefont {Wang},
  \citenamefont {Song}, \citenamefont {Sanchez}, \citenamefont {Yao},
  \citenamefont {Xiao}, \citenamefont {Zhu}, \citenamefont {May}, \citenamefont
  {Wu}, \citenamefont {Cobden}, \citenamefont {Chu},\ and\ \citenamefont
  {Xu}}]{Fei2018}%
  \BibitemOpen
  \bibfield  {author} {\bibinfo {author} {\bibfnamefont {Z.}~\bibnamefont
  {Fei}}, \bibinfo {author} {\bibfnamefont {B.}~\bibnamefont {Huang}}, \bibinfo
  {author} {\bibfnamefont {P.}~\bibnamefont {Malinowski}}, \bibinfo {author}
  {\bibfnamefont {W.}~\bibnamefont {Wang}}, \bibinfo {author} {\bibfnamefont
  {T.}~\bibnamefont {Song}}, \bibinfo {author} {\bibfnamefont {J.}~\bibnamefont
  {Sanchez}}, \bibinfo {author} {\bibfnamefont {W.}~\bibnamefont {Yao}},
  \bibinfo {author} {\bibfnamefont {D.}~\bibnamefont {Xiao}}, \bibinfo {author}
  {\bibfnamefont {X.}~\bibnamefont {Zhu}}, \bibinfo {author} {\bibfnamefont
  {A.~F.}\ \bibnamefont {May}}, \bibinfo {author} {\bibfnamefont
  {W.}~\bibnamefont {Wu}}, \bibinfo {author} {\bibfnamefont {D.~H.}\
  \bibnamefont {Cobden}}, \bibinfo {author} {\bibfnamefont {J.-H.}\
  \bibnamefont {Chu}}, \ and\ \bibinfo {author} {\bibfnamefont
  {X.}~\bibnamefont {Xu}},\ }\bibfield  {title} {\enquote {\bibinfo {title}
  {{Two-dimensional itinerant ferromagnetism in atomically thin
  \ce{Fe3GeTe2}}},}\ }\href {\doibase 10.1038/s41563-018-0149-7} {\bibfield
  {journal} {\bibinfo  {journal} {Nat. Mater.}\ }\textbf {\bibinfo {volume}
  {17}},\ \bibinfo {pages} {778--782} (\bibinfo {year} {2018})}\BibitemShut
  {NoStop}%
\bibitem [{\citenamefont {Deng}\ \emph {et~al.}(2018)\citenamefont {Deng},
  \citenamefont {Yu}, \citenamefont {Song}, \citenamefont {Zhang},
  \citenamefont {Wang}, \citenamefont {Sun}, \citenamefont {Yi}, \citenamefont
  {Wu}, \citenamefont {Wu}, \citenamefont {Zhu}, \citenamefont {Wang},
  \citenamefont {Chen},\ and\ \citenamefont {Zhang}}]{Deng2018a}%
  \BibitemOpen
  \bibfield  {author} {\bibinfo {author} {\bibfnamefont {Y.}~\bibnamefont
  {Deng}}, \bibinfo {author} {\bibfnamefont {Y.}~\bibnamefont {Yu}}, \bibinfo
  {author} {\bibfnamefont {Y.}~\bibnamefont {Song}}, \bibinfo {author}
  {\bibfnamefont {J.}~\bibnamefont {Zhang}}, \bibinfo {author} {\bibfnamefont
  {N.~Z.}\ \bibnamefont {Wang}}, \bibinfo {author} {\bibfnamefont
  {Z.}~\bibnamefont {Sun}}, \bibinfo {author} {\bibfnamefont {Y.}~\bibnamefont
  {Yi}}, \bibinfo {author} {\bibfnamefont {Y.~Z.}\ \bibnamefont {Wu}}, \bibinfo
  {author} {\bibfnamefont {S.}~\bibnamefont {Wu}}, \bibinfo {author}
  {\bibfnamefont {J.}~\bibnamefont {Zhu}}, \bibinfo {author} {\bibfnamefont
  {J.}~\bibnamefont {Wang}}, \bibinfo {author} {\bibfnamefont {X.~H.}\
  \bibnamefont {Chen}}, \ and\ \bibinfo {author} {\bibfnamefont
  {Y.}~\bibnamefont {Zhang}},\ }\bibfield  {title} {\enquote {\bibinfo {title}
  {{Gate-tunable room-temperature ferromagnetism in two-dimensional
  \ce{Fe3GeTe2}}},}\ }\href {\doibase 10.1038/s41586-018-0626-9} {\bibfield
  {journal} {\bibinfo  {journal} {Nature}\ }\textbf {\bibinfo {volume} {563}},\
  \bibinfo {pages} {94--99} (\bibinfo {year} {2018})}\BibitemShut {NoStop}%
\bibitem [{\citenamefont {Lai}\ \emph {et~al.}()\citenamefont {Lai},
  \citenamefont {Song}, \citenamefont {Wan}, \citenamefont {Xue}, \citenamefont
  {Ye}, \citenamefont {Dai}, \citenamefont {Yang}, \citenamefont {Du},\ and\
  \citenamefont {Yang}}]{Lai}%
  \BibitemOpen
  \bibfield  {author} {\bibinfo {author} {\bibfnamefont {Y.}~\bibnamefont
  {Lai}}, \bibinfo {author} {\bibfnamefont {Z.}~\bibnamefont {Song}}, \bibinfo
  {author} {\bibfnamefont {Y.}~\bibnamefont {Wan}}, \bibinfo {author}
  {\bibfnamefont {M.}~\bibnamefont {Xue}}, \bibinfo {author} {\bibfnamefont
  {Y.}~\bibnamefont {Ye}}, \bibinfo {author} {\bibfnamefont {L.}~\bibnamefont
  {Dai}}, \bibinfo {author} {\bibfnamefont {W.}~\bibnamefont {Yang}}, \bibinfo
  {author} {\bibfnamefont {H.}~\bibnamefont {Du}}, \ and\ \bibinfo {author}
  {\bibfnamefont {J.}~\bibnamefont {Yang}},\ }\bibfield  {title} {\enquote
  {\bibinfo {title} {{Discovery of two-dimensional multiferroicity in van der
  Waals \ce{CuCrP2S6} layers}},}\ }\href {http://arxiv.org/abs/1805.04280} {\
  }\Eprint {http://arxiv.org/abs/1805.04280} {arXiv:1805.04280} \BibitemShut
  {NoStop}%
\bibitem [{\citenamefont {Huang}\ \emph {et~al.}(2018)\citenamefont {Huang},
  \citenamefont {Clark}, \citenamefont {Klein}, \citenamefont {MacNeill},
  \citenamefont {Navarro-Moratalla}, \citenamefont {Seyler}, \citenamefont
  {Wilson}, \citenamefont {McGuire}, \citenamefont {Cobden}, \citenamefont
  {Xiao}, \citenamefont {Yao}, \citenamefont {Jarillo-Herrero},\ and\
  \citenamefont {Xu}}]{Huang}%
  \BibitemOpen
  \bibfield  {author} {\bibinfo {author} {\bibfnamefont {B.}~\bibnamefont
  {Huang}}, \bibinfo {author} {\bibfnamefont {G.}~\bibnamefont {Clark}},
  \bibinfo {author} {\bibfnamefont {D.~R.}\ \bibnamefont {Klein}}, \bibinfo
  {author} {\bibfnamefont {D.}~\bibnamefont {MacNeill}}, \bibinfo {author}
  {\bibfnamefont {E.}~\bibnamefont {Navarro-Moratalla}}, \bibinfo {author}
  {\bibfnamefont {K.~L.}\ \bibnamefont {Seyler}}, \bibinfo {author}
  {\bibfnamefont {N.}~\bibnamefont {Wilson}}, \bibinfo {author} {\bibfnamefont
  {M.~A.}\ \bibnamefont {McGuire}}, \bibinfo {author} {\bibfnamefont {D.~H.}\
  \bibnamefont {Cobden}}, \bibinfo {author} {\bibfnamefont {D.}~\bibnamefont
  {Xiao}}, \bibinfo {author} {\bibfnamefont {W.}~\bibnamefont {Yao}}, \bibinfo
  {author} {\bibfnamefont {P.}~\bibnamefont {Jarillo-Herrero}}, \ and\ \bibinfo
  {author} {\bibfnamefont {X.}~\bibnamefont {Xu}},\ }\bibfield  {title}
  {\enquote {\bibinfo {title} {{Electrical control of 2D magnetism in bilayer
  \ce{CrI3}}},}\ }\href {\doibase 10.1038/s41565-018-0121-3} {\bibfield
  {journal} {\bibinfo  {journal} {Nat. Nanotechnol.}\ }\textbf {\bibinfo
  {volume} {13}},\ \bibinfo {pages} {544--548} (\bibinfo {year}
  {2018})}\BibitemShut {NoStop}%
\bibitem [{\citenamefont {Jiang}\ \emph {et~al.}(2018)\citenamefont {Jiang},
  \citenamefont {Li}, \citenamefont {Wang}, \citenamefont {Mak},\ and\
  \citenamefont {Shan}}]{Jiang}%
  \BibitemOpen
  \bibfield  {author} {\bibinfo {author} {\bibfnamefont {S.}~\bibnamefont
  {Jiang}}, \bibinfo {author} {\bibfnamefont {L.}~\bibnamefont {Li}}, \bibinfo
  {author} {\bibfnamefont {Z.}~\bibnamefont {Wang}}, \bibinfo {author}
  {\bibfnamefont {K.~F.}\ \bibnamefont {Mak}}, \ and\ \bibinfo {author}
  {\bibfnamefont {J.}~\bibnamefont {Shan}},\ }\bibfield  {title} {\enquote
  {\bibinfo {title} {Controlling magnetism in 2d \ce{CrI3} by electrostatic
  doping},}\ }\href {\doibase 10.1038/s41565-018-0135-x} {\bibfield  {journal}
  {\bibinfo  {journal} {Nat. Nanotechnol.}\ }\textbf {\bibinfo {volume} {13}},\
  \bibinfo {pages} {549--553} (\bibinfo {year} {2018})}\BibitemShut {NoStop}%
\bibitem [{\citenamefont {Wang}\ \emph {et~al.}(2018)\citenamefont {Wang},
  \citenamefont {Zhang}, \citenamefont {Ding}, \citenamefont {Dong},
  \citenamefont {Li}, \citenamefont {Chen}, \citenamefont {Li}, \citenamefont
  {Huang}, \citenamefont {Wang}, \citenamefont {Zhao}, \citenamefont {Li},
  \citenamefont {Li}, \citenamefont {Jia}, \citenamefont {Sun}, \citenamefont
  {Guo}, \citenamefont {Ye}, \citenamefont {Sun}, \citenamefont {Chen},
  \citenamefont {Yang}, \citenamefont {Zhang}, \citenamefont {Ono},
  \citenamefont {Han},\ and\ \citenamefont {Zhang}}]{Wang}%
  \BibitemOpen
  \bibfield  {author} {\bibinfo {author} {\bibfnamefont {Z.}~\bibnamefont
  {Wang}}, \bibinfo {author} {\bibfnamefont {T.}~\bibnamefont {Zhang}},
  \bibinfo {author} {\bibfnamefont {M.}~\bibnamefont {Ding}}, \bibinfo {author}
  {\bibfnamefont {B.}~\bibnamefont {Dong}}, \bibinfo {author} {\bibfnamefont
  {Y.}~\bibnamefont {Li}}, \bibinfo {author} {\bibfnamefont {M.}~\bibnamefont
  {Chen}}, \bibinfo {author} {\bibfnamefont {X.}~\bibnamefont {Li}}, \bibinfo
  {author} {\bibfnamefont {J.}~\bibnamefont {Huang}}, \bibinfo {author}
  {\bibfnamefont {H.}~\bibnamefont {Wang}}, \bibinfo {author} {\bibfnamefont
  {X.}~\bibnamefont {Zhao}}, \bibinfo {author} {\bibfnamefont {Y.}~\bibnamefont
  {Li}}, \bibinfo {author} {\bibfnamefont {D.}~\bibnamefont {Li}}, \bibinfo
  {author} {\bibfnamefont {C.}~\bibnamefont {Jia}}, \bibinfo {author}
  {\bibfnamefont {L.}~\bibnamefont {Sun}}, \bibinfo {author} {\bibfnamefont
  {H.}~\bibnamefont {Guo}}, \bibinfo {author} {\bibfnamefont {Yu}~\bibnamefont
  {Ye}}, \bibinfo {author} {\bibfnamefont {D.}~\bibnamefont {Sun}}, \bibinfo
  {author} {\bibfnamefont {Y.}~\bibnamefont {Chen}}, \bibinfo {author}
  {\bibfnamefont {T.}~\bibnamefont {Yang}}, \bibinfo {author} {\bibfnamefont
  {J.}~\bibnamefont {Zhang}}, \bibinfo {author} {\bibfnamefont
  {S.}~\bibnamefont {Ono}}, \bibinfo {author} {\bibfnamefont {Z.}~\bibnamefont
  {Han}}, \ and\ \bibinfo {author} {\bibfnamefont {Z.}~\bibnamefont {Zhang}},\
  }\bibfield  {title} {\enquote {\bibinfo {title} {Electric-field control of
  magnetism in a few-layered van der waals ferromagnetic semiconductor},}\
  }\href {\doibase 10.1038/s41565-018-0186-z} {\bibfield  {journal} {\bibinfo
  {journal} {Nat. Nanotechnol.}\ }\textbf {\bibinfo {volume} {13}},\ \bibinfo
  {pages} {554--559} (\bibinfo {year} {2018})}\BibitemShut {NoStop}%
\bibitem [{\citenamefont {Kim}\ \emph {et~al.}(2018)\citenamefont {Kim},
  \citenamefont {Yang}, \citenamefont {Patel}, \citenamefont {Sfigakis},
  \citenamefont {Li}, \citenamefont {Tian}, \citenamefont {Lei},\ and\
  \citenamefont {Tsen}}]{Kim}%
  \BibitemOpen
  \bibfield  {author} {\bibinfo {author} {\bibfnamefont {H.~H.}\ \bibnamefont
  {Kim}}, \bibinfo {author} {\bibfnamefont {B.}~\bibnamefont {Yang}}, \bibinfo
  {author} {\bibfnamefont {T.}~\bibnamefont {Patel}}, \bibinfo {author}
  {\bibfnamefont {F.}~\bibnamefont {Sfigakis}}, \bibinfo {author}
  {\bibfnamefont {C.}~\bibnamefont {Li}}, \bibinfo {author} {\bibfnamefont
  {S.}~\bibnamefont {Tian}}, \bibinfo {author} {\bibfnamefont {H.}~\bibnamefont
  {Lei}}, \ and\ \bibinfo {author} {\bibfnamefont {A.~W.}\ \bibnamefont
  {Tsen}},\ }\bibfield  {title} {\enquote {\bibinfo {title} {{One million
  percent tunnel magnetoresistance in a magnetic van der Waals
  heterostructure}},}\ }\href {\doibase 10.1021/acs.nanolett.8b01552}
  {\bibfield  {journal} {\bibinfo  {journal} {Nano Lett.}\ }\textbf {\bibinfo
  {volume} {18}},\ \bibinfo {pages} {4885--4890} (\bibinfo {year}
  {2018})}\BibitemShut {NoStop}%
\bibitem [{\citenamefont {Brataas}\ \emph {et~al.}(2012)\citenamefont
  {Brataas}, \citenamefont {Kent},\ and\ \citenamefont {Ohno}}]{Brataas2012}%
  \BibitemOpen
  \bibfield  {author} {\bibinfo {author} {\bibfnamefont {A.}~\bibnamefont
  {Brataas}}, \bibinfo {author} {\bibfnamefont {A.~D.}\ \bibnamefont {Kent}}, \
  and\ \bibinfo {author} {\bibfnamefont {H.}~\bibnamefont {Ohno}},\ }\bibfield
  {title} {\enquote {\bibinfo {title} {Current-induced torques in magnetic
  materials},}\ }\href {\doibase 10.1038/nmat3311} {\bibfield  {journal}
  {\bibinfo  {journal} {Nat. Mater.}\ }\textbf {\bibinfo {volume} {11}},\
  \bibinfo {pages} {372} (\bibinfo {year} {2012})}\BibitemShut {NoStop}%
\bibitem [{\citenamefont {Manchon}\ and\ \citenamefont
  {Zhang}(2008)}]{Manchon2008}%
  \BibitemOpen
  \bibfield  {author} {\bibinfo {author} {\bibfnamefont {A.}~\bibnamefont
  {Manchon}}\ and\ \bibinfo {author} {\bibfnamefont {S.}~\bibnamefont
  {Zhang}},\ }\bibfield  {title} {\enquote {\bibinfo {title} {{Theory of
  nonequilibrium intrinsic spin torque in a single nanomagnet}},}\ }\href
  {\doibase 10.1103/PhysRevB.78.212405} {\bibfield  {journal} {\bibinfo
  {journal} {Phys. Rev. B}\ }\textbf {\bibinfo {volume} {78}},\ \bibinfo
  {pages} {212405} (\bibinfo {year} {2008})}\BibitemShut {NoStop}%
\bibitem [{\citenamefont {Manchon}\ \emph {et~al.}()\citenamefont {Manchon},
  \citenamefont {Miron}, \citenamefont {Jungwirth}, \citenamefont {Sinova},
  \citenamefont {Zelenzný}, \citenamefont {Thiaville}, \citenamefont
  {Garello},\ and\ \citenamefont {Gambardella}}]{Manchon}%
  \BibitemOpen
  \bibfield  {author} {\bibinfo {author} {\bibfnamefont {A.}~\bibnamefont
  {Manchon}}, \bibinfo {author} {\bibfnamefont {I.~M.}\ \bibnamefont {Miron}},
  \bibinfo {author} {\bibfnamefont {T.}~\bibnamefont {Jungwirth}}, \bibinfo
  {author} {\bibfnamefont {J.}~\bibnamefont {Sinova}}, \bibinfo {author}
  {\bibfnamefont {J.}~\bibnamefont {Zelenzný}}, \bibinfo {author}
  {\bibfnamefont {A.}~\bibnamefont {Thiaville}}, \bibinfo {author}
  {\bibfnamefont {K.}~\bibnamefont {Garello}}, \ and\ \bibinfo {author}
  {\bibfnamefont {P.}~\bibnamefont {Gambardella}},\ }\bibfield  {title}
  {\enquote {\bibinfo {title} {{Current-induced spin-orbit torques in
  ferromagnetic and antiferromagnetic systems}},}\ }\href
  {http://arxiv.org/abs/1801.09636} {\ }\Eprint
  {http://arxiv.org/abs/1801.09636} {arXiv:1801.09636} \BibitemShut {NoStop}%
\bibitem [{\citenamefont {Kaplan}\ \emph {et~al.}(2009)\citenamefont {Kaplan},
  \citenamefont {Lee}, \citenamefont {Son},\ and\ \citenamefont
  {Stephanov}}]{Kaplan2009conformality}%
  \BibitemOpen
  \bibfield  {author} {\bibinfo {author} {\bibfnamefont {D.~B.}\ \bibnamefont
  {Kaplan}}, \bibinfo {author} {\bibfnamefont {J.-W.}\ \bibnamefont {Lee}},
  \bibinfo {author} {\bibfnamefont {D.~T.}\ \bibnamefont {Son}}, \ and\
  \bibinfo {author} {\bibfnamefont {M.~A.}\ \bibnamefont {Stephanov}},\
  }\bibfield  {title} {\enquote {\bibinfo {title} {Conformality lost},}\ }\href
  {\doibase 10.1103/PhysRevD.80.125005} {\bibfield  {journal} {\bibinfo
  {journal} {Phys. Rev. D}\ }\textbf {\bibinfo {volume} {80}},\ \bibinfo
  {pages} {125005} (\bibinfo {year} {2009})}\BibitemShut {NoStop}%
\bibitem [{\citenamefont {Deiseroth}\ \emph {et~al.}(2006)\citenamefont
  {Deiseroth}, \citenamefont {Aleksandrov}, \citenamefont {Reiner},
  \citenamefont {Kienle},\ and\ \citenamefont {Kremer}}]{Deiseroth2006}%
  \BibitemOpen
  \bibfield  {author} {\bibinfo {author} {\bibfnamefont {H.-J.}\ \bibnamefont
  {Deiseroth}}, \bibinfo {author} {\bibfnamefont {K.}~\bibnamefont
  {Aleksandrov}}, \bibinfo {author} {\bibfnamefont {C.}~\bibnamefont {Reiner}},
  \bibinfo {author} {\bibfnamefont {L.}~\bibnamefont {Kienle}}, \ and\ \bibinfo
  {author} {\bibfnamefont {R.~K.}\ \bibnamefont {Kremer}},\ }\bibfield  {title}
  {\enquote {\bibinfo {title} {{\ce{Fe3GeTe2} and \ce{Ni3GeTe2} – two new
  layered transition-metal compounds: Crystal structures, HRTEM investigations,
  and magnetic and electrical properties}},}\ }\href {\doibase
  10.1002/ejic.200501020} {\bibfield  {journal} {\bibinfo  {journal} {Eur. J.
  Inorg. Chem.}\ }\textbf {\bibinfo {volume} {2006}},\ \bibinfo {pages}
  {1561--1567} (\bibinfo {year} {2006})}\BibitemShut {NoStop}%
\bibitem [{\citenamefont {Hals}\ and\ \citenamefont
  {Brataas}(2013)}]{Hals2013}%
  \BibitemOpen
  \bibfield  {author} {\bibinfo {author} {\bibfnamefont {K.~M.~D.}\
  \bibnamefont {Hals}}\ and\ \bibinfo {author} {\bibfnamefont {A.}~\bibnamefont
  {Brataas}},\ }\bibfield  {title} {\enquote {\bibinfo {title} {{Phenomenology
  of current-induced spin-orbit torques}},}\ }\href {\doibase
  10.1103/PhysRevB.88.085423} {\bibfield  {journal} {\bibinfo  {journal} {Phys.
  Rev. B}\ }\textbf {\bibinfo {volume} {88}},\ \bibinfo {pages} {085423}
  (\bibinfo {year} {2013})}\BibitemShut {NoStop}%
\bibitem [{\citenamefont {Hals}\ and\ \citenamefont
  {Brataas}(2015)}]{Hals2015}%
  \BibitemOpen
  \bibfield  {author} {\bibinfo {author} {\bibfnamefont {K.~M.~D.}\
  \bibnamefont {Hals}}\ and\ \bibinfo {author} {\bibfnamefont {A.}~\bibnamefont
  {Brataas}},\ }\bibfield  {title} {\enquote {\bibinfo {title} {{Spin-motive
  forces and current-induced torques in ferromagnets}},}\ }\href {\doibase
  10.1103/PhysRevB.91.214401} {\bibfield  {journal} {\bibinfo  {journal} {Phys.
  Rev. B}\ }\textbf {\bibinfo {volume} {91}},\ \bibinfo {pages} {214401}
  (\bibinfo {year} {2015})}\BibitemShut {NoStop}%
\bibitem [{Note1()}]{Note1}%
  \BibitemOpen
  \bibinfo {note} {See the Supplemental Material, which includes Ref.~\cite
  {Birss1964}, a derivation of the spin--orbit torques, and the details of the
  critical-temperature calculation.}\BibitemShut {Stop}%
\bibitem [{\citenamefont {Zhuang}\ \emph {et~al.}(2016)\citenamefont {Zhuang},
  \citenamefont {Kent},\ and\ \citenamefont {Hennig}}]{Zhuang:prb:2016}%
  \BibitemOpen
  \bibfield  {author} {\bibinfo {author} {\bibfnamefont {H.~L.}\ \bibnamefont
  {Zhuang}}, \bibinfo {author} {\bibfnamefont {P.~R.~C.}\ \bibnamefont {Kent}},
  \ and\ \bibinfo {author} {\bibfnamefont {R.~G.}\ \bibnamefont {Hennig}},\
  }\bibfield  {title} {\enquote {\bibinfo {title} {Strong anisotropy and
  magnetostriction in the two-dimensional {Stoner} ferromagnet
  \ce{Fe3GeTe2}},}\ }\href {\doibase 10.1103/PhysRevB.93.134407} {\bibfield
  {journal} {\bibinfo  {journal} {Phys. Rev. B}\ }\textbf {\bibinfo {volume}
  {93}},\ \bibinfo {pages} {134407} (\bibinfo {year} {2016})}\BibitemShut
  {NoStop}%
\bibitem [{\citenamefont {Chartoryzhskii}\ \emph {et~al.}(1976)\citenamefont
  {Chartoryzhskii}, \citenamefont {Kalinikos},\ and\ \citenamefont
  {Vendik}}]{Chartoryzhskii1976}%
  \BibitemOpen
  \bibfield  {author} {\bibinfo {author} {\bibfnamefont {D.~N.}\ \bibnamefont
  {Chartoryzhskii}}, \bibinfo {author} {\bibfnamefont {B.~A.}\ \bibnamefont
  {Kalinikos}}, \ and\ \bibinfo {author} {\bibfnamefont {O.~G.}\ \bibnamefont
  {Vendik}},\ }\bibfield  {title} {\enquote {\bibinfo {title} {{Parallel pump
  spin wave instability in thin ferromagnetic films}},}\ }\href {\doibase
  10.1016/0038-1098(76)90489-0} {\bibfield  {journal} {\bibinfo  {journal}
  {Solid State Commun.}\ }\textbf {\bibinfo {volume} {20}},\ \bibinfo {pages}
  {985--989} (\bibinfo {year} {1976})}\BibitemShut {NoStop}%
\bibitem [{\citenamefont {Kalinikos}(1980)}]{Kalinikos1980}%
  \BibitemOpen
  \bibfield  {author} {\bibinfo {author} {\bibfnamefont {B.~A.}\ \bibnamefont
  {Kalinikos}},\ }\bibfield  {title} {\enquote {\bibinfo {title} {{Excitation
  of propagating spin waves in ferromagnetic films}},}\ }\href {\doibase
  10.1049/ip-h-1.1980.0002} {\bibfield  {journal} {\bibinfo  {journal} {IEE
  Proceedings H}\ }\textbf {\bibinfo {volume} {127}},\ \bibinfo {pages} {4}
  (\bibinfo {year} {1980})}\BibitemShut {NoStop}%
\bibitem [{\citenamefont {Kalinikos}(1981)}]{Kalinikos1981a}%
  \BibitemOpen
  \bibfield  {author} {\bibinfo {author} {\bibfnamefont {B.~A.}\ \bibnamefont
  {Kalinikos}},\ }\bibfield  {title} {\enquote {\bibinfo {title} {{Spectrum and
  linear excitation of spin waves in ferromagnetic films}},}\ }\href {\doibase
  10.1007/BF00941342} {\bibfield  {journal} {\bibinfo  {journal} {Soviet
  Physics Journal}\ }\textbf {\bibinfo {volume} {24}},\ \bibinfo {pages}
  {718--731} (\bibinfo {year} {1981})}\BibitemShut {NoStop}%
\bibitem [{\citenamefont {Kalinikos}\ and\ \citenamefont
  {Slavin}(1986)}]{Kalinikos1986}%
  \BibitemOpen
  \bibfield  {author} {\bibinfo {author} {\bibfnamefont {B.~A.}\ \bibnamefont
  {Kalinikos}}\ and\ \bibinfo {author} {\bibfnamefont {A.~N.}\ \bibnamefont
  {Slavin}},\ }\bibfield  {title} {\enquote {\bibinfo {title} {{Theory of
  dipole-exchange spin wave spectrum for ferromagnetic films with mixed
  exchange boundary conditions}},}\ }\href {\doibase
  10.1088/0022-3719/19/35/014} {\bibfield  {journal} {\bibinfo  {journal} {J.
  Phys. C: Solid State}\ }\textbf {\bibinfo {volume} {19}},\ \bibinfo {pages}
  {7013--7033} (\bibinfo {year} {1986})}\BibitemShut {NoStop}%
\bibitem [{\citenamefont {Kalinikos}\ \emph {et~al.}(1990)\citenamefont
  {Kalinikos}, \citenamefont {Kostylev}, \citenamefont {Kozhus},\ and\
  \citenamefont {Slavin}}]{Kalinikos1990}%
  \BibitemOpen
  \bibfield  {author} {\bibinfo {author} {\bibfnamefont {B.~A.}\ \bibnamefont
  {Kalinikos}}, \bibinfo {author} {\bibfnamefont {M.~P.}\ \bibnamefont
  {Kostylev}}, \bibinfo {author} {\bibfnamefont {N.~V.}\ \bibnamefont
  {Kozhus}}, \ and\ \bibinfo {author} {\bibfnamefont {A.~N.}\ \bibnamefont
  {Slavin}},\ }\bibfield  {title} {\enquote {\bibinfo {title} {{The
  dipole-exchange spin wave spectrum for anisotropic ferromagnetic films with
  mixed exchange boundary conditions}},}\ }\href {\doibase
  10.1088/0953-8984/2/49/012} {\bibfield  {journal} {\bibinfo  {journal} {J.
  Phys. Condens. Matt.}\ }\textbf {\bibinfo {volume} {2}},\ \bibinfo {pages}
  {9861--9877} (\bibinfo {year} {1990})}\BibitemShut {NoStop}%
\bibitem [{\citenamefont {Auerbach}(1994)}]{Auerbach1994}%
  \BibitemOpen
  \bibfield  {author} {\bibinfo {author} {\bibfnamefont {A.}~\bibnamefont
  {Auerbach}},\ }\href@noop {} {\emph {\bibinfo {title} {{Interacting Electrons
  and Quantum Magnetism}}}},\ Graduate Texts in Contemporary Physics\ (\bibinfo
   {publisher} {Springer-Verlag},\ \bibinfo {year} {1994})\BibitemShut
  {NoStop}%
\bibitem [{\citenamefont {Prange}\ and\ \citenamefont
  {Korenman}(1979)}]{Prange:prb:1979}%
  \BibitemOpen
  \bibfield  {author} {\bibinfo {author} {\bibfnamefont {R.~E.}\ \bibnamefont
  {Prange}}\ and\ \bibinfo {author} {\bibfnamefont {V.}~\bibnamefont
  {Korenman}},\ }\bibfield  {title} {\enquote {\bibinfo {title} {Local-band
  theory of itinerant ferromagnetism. {IV.} {Equivalent} {Heisenberg} model},}\
  }\href {\doibase 10.1103/PhysRevB.19.4691} {\bibfield  {journal} {\bibinfo
  {journal} {Phys. Rev. B}\ }\textbf {\bibinfo {volume} {19}},\ \bibinfo
  {pages} {4691--4697} (\bibinfo {year} {1979})}\BibitemShut {NoStop}%
\bibitem [{\citenamefont {Tan}\ \emph {et~al.}(2018)\citenamefont {Tan},
  \citenamefont {Lee}, \citenamefont {Jung}, \citenamefont {Park},
  \citenamefont {Albarakati}, \citenamefont {Partridge}, \citenamefont {Field},
  \citenamefont {McCulloch}, \citenamefont {Wang},\ and\ \citenamefont
  {Lee}}]{Tan2018}%
  \BibitemOpen
  \bibfield  {author} {\bibinfo {author} {\bibfnamefont {C.}~\bibnamefont
  {Tan}}, \bibinfo {author} {\bibfnamefont {J.}~\bibnamefont {Lee}}, \bibinfo
  {author} {\bibfnamefont {S.-G.}\ \bibnamefont {Jung}}, \bibinfo {author}
  {\bibfnamefont {T.}~\bibnamefont {Park}}, \bibinfo {author} {\bibfnamefont
  {S.}~\bibnamefont {Albarakati}}, \bibinfo {author} {\bibfnamefont
  {J.}~\bibnamefont {Partridge}}, \bibinfo {author} {\bibfnamefont {M.~R.}\
  \bibnamefont {Field}}, \bibinfo {author} {\bibfnamefont {D.~G.}\ \bibnamefont
  {McCulloch}}, \bibinfo {author} {\bibfnamefont {L.}~\bibnamefont {Wang}}, \
  and\ \bibinfo {author} {\bibfnamefont {C.}~\bibnamefont {Lee}},\ }\bibfield
  {title} {\enquote {\bibinfo {title} {Hard magnetic properties in nanoflake
  van der {W}aals \ce{Fe3GeTe2}},}\ }\href {\doibase
  10.1038/s41467-018-04018-w} {\bibfield  {journal} {\bibinfo  {journal} {Nat.
  Commun.}\ }\textbf {\bibinfo {volume} {9}},\ \bibinfo {pages} {1554}
  (\bibinfo {year} {2018})}\BibitemShut {NoStop}%
\bibitem [{\citenamefont {Nagaosa}\ \emph {et~al.}(2010)\citenamefont
  {Nagaosa}, \citenamefont {Sinova}, \citenamefont {Onoda}, \citenamefont
  {MacDonald},\ and\ \citenamefont {Ong}}]{Nagaosa2010}%
  \BibitemOpen
  \bibfield  {author} {\bibinfo {author} {\bibfnamefont {N.}~\bibnamefont
  {Nagaosa}}, \bibinfo {author} {\bibfnamefont {J.}~\bibnamefont {Sinova}},
  \bibinfo {author} {\bibfnamefont {S.}~\bibnamefont {Onoda}}, \bibinfo
  {author} {\bibfnamefont {A.~H.}\ \bibnamefont {MacDonald}}, \ and\ \bibinfo
  {author} {\bibfnamefont {N.~P.}\ \bibnamefont {Ong}},\ }\bibfield  {title}
  {\enquote {\bibinfo {title} {{Anomalous Hall effect}},}\ }\href {\doibase
  10.1103/RevModPhys.82.1539} {\bibfield  {journal} {\bibinfo  {journal} {Rev.
  Mod. Phys.}\ }\textbf {\bibinfo {volume} {82}},\ \bibinfo {pages} {1539}
  (\bibinfo {year} {2010})}\BibitemShut {NoStop}%
\bibitem [{\citenamefont {Kamra}\ \emph {et~al.}(2017)\citenamefont {Kamra},
  \citenamefont {Agrawal},\ and\ \citenamefont {Belzig}}]{Kamra:prb:2017}%
  \BibitemOpen
  \bibfield  {author} {\bibinfo {author} {\bibfnamefont {A.}~\bibnamefont
  {Kamra}}, \bibinfo {author} {\bibfnamefont {U.}~\bibnamefont {Agrawal}}, \
  and\ \bibinfo {author} {\bibfnamefont {W.}~\bibnamefont {Belzig}},\
  }\bibfield  {title} {\enquote {\bibinfo {title} {Noninteger-spin magnonic
  excitations in untextured magnets},}\ }\href {\doibase
  10.1103/PhysRevB.96.020411} {\bibfield  {journal} {\bibinfo  {journal} {Phys.
  Rev. B}\ }\textbf {\bibinfo {volume} {96}},\ \bibinfo {pages} {020411}
  (\bibinfo {year} {2017})}\BibitemShut {NoStop}%
\bibitem [{\citenamefont {Rodriguez}\ \emph {et~al.}(1996)\citenamefont
  {Rodriguez}, \citenamefont {Ok}, \citenamefont {Xia}, \citenamefont
  {Debrunner}, \citenamefont {Hinrichs}, \citenamefont {Meyer},\ and\
  \citenamefont {Packard}}]{Rodriguez:JPC:1996}%
  \BibitemOpen
  \bibfield  {author} {\bibinfo {author} {\bibfnamefont {J.~H.}\ \bibnamefont
  {Rodriguez}}, \bibinfo {author} {\bibfnamefont {H.~N.}\ \bibnamefont {Ok}},
  \bibinfo {author} {\bibfnamefont {Y.-M.}\ \bibnamefont {Xia}}, \bibinfo
  {author} {\bibfnamefont {P.~G.}\ \bibnamefont {Debrunner}}, \bibinfo {author}
  {\bibfnamefont {B.~E.}\ \bibnamefont {Hinrichs}}, \bibinfo {author}
  {\bibfnamefont {T.}~\bibnamefont {Meyer}}, \ and\ \bibinfo {author}
  {\bibfnamefont {N.~H.}\ \bibnamefont {Packard}},\ }\bibfield  {title}
  {\enquote {\bibinfo {title} {M\"{o}ssbauer spectroscopy of the spin-coupled
  $\mathrm{{F}e^{3+}-{F}e^{2+}}$ center of reduced uteroferrin},}\ }\href
  {\doibase 10.1021/jp9532529} {\bibfield  {journal} {\bibinfo  {journal} {J.
  Phys. Chem.}\ }\textbf {\bibinfo {volume} {100}},\ \bibinfo {pages}
  {6849--6862} (\bibinfo {year} {1996})}\BibitemShut {NoStop}%
\bibitem [{\citenamefont {Torelli}\ and\ \citenamefont {Olsen}()}]{Torelli}%
  \BibitemOpen
  \bibfield  {author} {\bibinfo {author} {\bibfnamefont {D.}~\bibnamefont
  {Torelli}}\ and\ \bibinfo {author} {\bibfnamefont {T.}~\bibnamefont
  {Olsen}},\ }\bibfield  {title} {\enquote {\bibinfo {title} {{Calculating
  critical temperatures for ferromagnetic order in two-dimensional
  materials}},}\ }\href {http://arxiv.org/abs/1808.06400} {\ }\Eprint
  {http://arxiv.org/abs/1808.06400} {arXiv:1808.06400} \BibitemShut {NoStop}%
\bibitem [{\citenamefont {Landau}\ \emph {et~al.}(1999)\citenamefont {Landau},
  \citenamefont {Lifshitz},\ and\ \citenamefont
  {Pitaevskii}}]{Landau-Ginzburg}%
  \BibitemOpen
  \bibfield  {author} {\bibinfo {author} {\bibfnamefont {L.~D.}\ \bibnamefont
  {Landau}}, \bibinfo {author} {\bibfnamefont {E.~M.}\ \bibnamefont
  {Lifshitz}}, \ and\ \bibinfo {author} {\bibfnamefont {E.~M.}\ \bibnamefont
  {Pitaevskii}},\ }\href@noop {} {\emph {\bibinfo {title} {{Statistical
  Physics}}}},\ \bibinfo {edition} {1st}\ ed.\ (\bibinfo  {publisher}
  {Buttwerworth and Heinemann},\ \bibinfo {year} {1999})\BibitemShut {NoStop}%
\bibitem [{\citenamefont {Wilson}\ and\ \citenamefont
  {Kogut}(1974)}]{Wilson-Kogut}%
  \BibitemOpen
  \bibfield  {author} {\bibinfo {author} {\bibfnamefont {K.~G.}\ \bibnamefont
  {Wilson}}\ and\ \bibinfo {author} {\bibfnamefont {J.~B}\ \bibnamefont
  {Kogut}},\ }\bibfield  {title} {\enquote {\bibinfo {title} {{The
  renormalization group and the $\varepsilon$ expansion}},}\ }\href {\doibase
  10.1049/ip-h-1.1980.0002} {\bibfield  {journal} {\bibinfo  {journal} {Phys.
  Rep}\ }\textbf {\bibinfo {volume} {12}},\ \bibinfo {pages} {74--200}
  (\bibinfo {year} {1974})}\BibitemShut {NoStop}%
\bibitem [{\citenamefont {Maurer}\ \emph {et~al.}(2014)\citenamefont {Maurer},
  \citenamefont {Gautrot}, \citenamefont {Ott}, \citenamefont {Chaboussant},
  \citenamefont {Zighem}, \citenamefont {Cagnon},\ and\ \citenamefont
  {Fruchart}}]{Maurer}%
  \BibitemOpen
  \bibfield  {author} {\bibinfo {author} {\bibfnamefont {T.}~\bibnamefont
  {Maurer}}, \bibinfo {author} {\bibfnamefont {S.}~\bibnamefont {Gautrot}},
  \bibinfo {author} {\bibfnamefont {F.}~\bibnamefont {Ott}}, \bibinfo {author}
  {\bibfnamefont {G.}~\bibnamefont {Chaboussant}}, \bibinfo {author}
  {\bibfnamefont {F.}~\bibnamefont {Zighem}}, \bibinfo {author} {\bibfnamefont
  {L.}~\bibnamefont {Cagnon}}, \ and\ \bibinfo {author} {\bibfnamefont
  {O.}~\bibnamefont {Fruchart}},\ }\bibfield  {title} {\enquote {\bibinfo
  {title} {Ordered arrays of magnetic nanowires investigated by polarized
  small-angle neutron scattering},}\ }\href {\doibase
  10.1103/PhysRevB.89.184423} {\bibfield  {journal} {\bibinfo  {journal} {Phys.
  Rev. B}\ }\textbf {\bibinfo {volume} {89}},\ \bibinfo {pages} {184423}
  (\bibinfo {year} {2014})}\BibitemShut {NoStop}%
\bibitem [{\citenamefont {Itzykson}\ and\ \citenamefont
  {Drouffe}(1989)}]{Itzykson-Drouffe}%
  \BibitemOpen
  \bibfield  {author} {\bibinfo {author} {\bibfnamefont {C.}~\bibnamefont
  {Itzykson}}\ and\ \bibinfo {author} {\bibfnamefont {J.-M.}\ \bibnamefont
  {Drouffe}},\ }\href@noop {} {\emph {\bibinfo {title} {{Statistical field
  theory}}}},\ \bibinfo {series} {Cambridge Monographs on Mathematical
  Physics}, Vol.~\bibinfo {volume} {1}\ (\bibinfo  {publisher} {Cambridge
  University Press},\ \bibinfo {year} {1989})\BibitemShut {NoStop}%
\bibitem [{\citenamefont {Bishop}\ and\ \citenamefont
  {Reppy}(1978)}]{BishopReppy}%
  \BibitemOpen
  \bibfield  {author} {\bibinfo {author} {\bibfnamefont {D.~J.}\ \bibnamefont
  {Bishop}}\ and\ \bibinfo {author} {\bibfnamefont {J.~D.}\ \bibnamefont
  {Reppy}},\ }\bibfield  {title} {\enquote {\bibinfo {title} {Study of the
  superfluid transition in two-dimensional $^{4}\mathrm{He}$ films},}\ }\href
  {\doibase 10.1103/PhysRevLett.40.1727} {\bibfield  {journal} {\bibinfo
  {journal} {Phys. Rev. Lett.}\ }\textbf {\bibinfo {volume} {40}},\ \bibinfo
  {pages} {1727--1730} (\bibinfo {year} {1978})}\BibitemShut {NoStop}%
\bibitem [{\citenamefont {Nelson}\ and\ \citenamefont
  {Kosterlitz}(1977)}]{nelson1977universal}%
  \BibitemOpen
  \bibfield  {author} {\bibinfo {author} {\bibfnamefont {D.~R.}\ \bibnamefont
  {Nelson}}\ and\ \bibinfo {author} {\bibfnamefont {J.~M.}\ \bibnamefont
  {Kosterlitz}},\ }\bibfield  {title} {\enquote {\bibinfo {title} {Universal
  jump in the superfluid density of two-dimensional superfluids},}\ }\href
  {\doibase 10.1103/PhysRevLett.39.1201} {\bibfield  {journal} {\bibinfo
  {journal} {Phys. Rev. Lett.}\ }\textbf {\bibinfo {volume} {39}},\ \bibinfo
  {pages} {1201--1205} (\bibinfo {year} {1977})}\BibitemShut {NoStop}%
\bibitem [{\citenamefont {Golosovsky}\ \emph {et~al.}(2007)\citenamefont
  {Golosovsky}, \citenamefont {Monod}, \citenamefont {Muduli},\ and\
  \citenamefont {Budhani}}]{Golosovsky}%
  \BibitemOpen
  \bibfield  {author} {\bibinfo {author} {\bibfnamefont {M.}~\bibnamefont
  {Golosovsky}}, \bibinfo {author} {\bibfnamefont {P.}~\bibnamefont {Monod}},
  \bibinfo {author} {\bibfnamefont {P.~K.}\ \bibnamefont {Muduli}}, \ and\
  \bibinfo {author} {\bibfnamefont {R.~C.}\ \bibnamefont {Budhani}},\
  }\bibfield  {title} {\enquote {\bibinfo {title} {Spin-wave resonances in
  \ce{La_{0.7}Sr_{0.3}MnO_3} films: Measurement of spin-wave stiffness and
  anisotropy field},}\ }\href {\doibase 10.1103/PhysRevB.76.184413} {\bibfield
  {journal} {\bibinfo  {journal} {Phys. Rev. B}\ }\textbf {\bibinfo {volume}
  {76}},\ \bibinfo {pages} {184413} (\bibinfo {year} {2007})}\BibitemShut
  {NoStop}%
\bibitem [{\citenamefont {Birss}(1964)}]{Birss1964}%
  \BibitemOpen
  \bibfield  {author} {\bibinfo {author} {\bibfnamefont {R.~R.}\ \bibnamefont
  {Birss}},\ }\href@noop {} {\emph {\bibinfo {title} {{Symmetry and
  Magnetism}}}},\ \bibinfo {edition} {1st}\ ed.,\ edited by\ \bibinfo {editor}
  {\bibfnamefont {E.~P.}\ \bibnamefont {Wohlfarth}},\ \bibinfo {series}
  {Selected Topics in Solid State Physics}, Vol.~\bibinfo {volume} {3}\
  (\bibinfo  {publisher} {North-Holland Publishing Company},\ \bibinfo {year}
  {1964})\BibitemShut {NoStop}%
\end{thebibliography}
\end{document}


\title{Supplementary Material to \\ ``Current Control of Magnetism in Two-Dimensional \ce{Fe3GeTe2}''}
\author{Øyvind Johansen}
\email{oyvinjoh@ntnu.no}
\author{Vetle Risinggård}
\email{vetle.k.risinggard@ntnu.no}
\author{Asle Sudbø}
\author{Jacob Linder}
\author{Arne Brataas}
\affiliation{Center for Quantum Spintronics, Department of Physics, NTNU, Norwegian University of Science and Technology, N-7491 Trondheim, Norway}
\date{\today}

\maketitle

\section{Derivation of spin--orbit torques in $\mathbf{\mathbf{Fe}_3\mathbf{GeTe}_2}$}
Ref.~\cite{Hals2013,*Hals2015} has shown that in the linear response regime under the local approximation, the current-induced torques can be written as
\begin{align}
&\vec\tau(\vec r,t)=-|\gamma|\vec m(\vec r,t)\times\vec{H}_\mathrm{SOT}, &&H_{\mathrm{SOT},i}=\eta_{ij\,}J_j\,,
\end{align}
where $\vec m$ is the magnetization unit vector, $\vec J$ is the current density applied to the system, and $\eta$ is a second-rank tensor. 
(Summation over repeated indices is implied.) 
Which elements of $\eta_{ij}$ that are nonzero is determined by the symmetry of the system. 
The tensor $\eta$ can be expanded in the magnetization components $m_i$ and their derivatives $\partial_im_j$. 
If we only consider a uniform magnetization, one obtains to lowest order
\begin{equation}
\eta_{ij}=\Lambda_{ij}+\Gamma_{ijk}m_k+\dots
\end{equation}
where $\Lambda_{ij}$ is an axial second-rank tensor, and $\Gamma_{ijk}$ is a polar third-rank tensor.

To determine which contributions to the tensors $\Lambda_{ij}$ and $\Gamma_{ijk}$ that are allowed by symmetry, Ref.~\cite{Hals2013} imposes the criterion that these tensors must be invariant under all transformations $R$ in the point group $G$ of the structure. 
This amounts to demanding that the relations
\begin{align}
&\Lambda_{ij}=|R|R_{ii'}R_{jj'}\Lambda_{i'j'}, \label{eq:reactive}\\
&\Gamma_{ijk}=R_{ii'}R_{jj'}R_{kk'}\Gamma_{i'j'k'}, \label{eq:dissipative}
\end{align}
are fulfilled for all $R\in G$.

Monolayer \ce{Fe3GeTe2} crystallizes in point group $\sym{\bar 6m2}$ ($D_{3h}$)~\cite{Deiseroth2006}. 
Since this group is generated by the elements $\symsub{\bar 6}{z}$, $\symsub{m}{y}$, and $\symsub{2}{x}$, it is sufficient to impose that $\eta_{ij}$ should be invariant under these operations~\cite{Birss1964}. 
The representing matrices of these symmetry operations are
\begin{align*}
&\symsub{m}{y}\!=\!\left(\!\!\begin{array}{rrr} 1 & & \\ & -1 & \\ & & 1 \end{array}\!\!\right), 
&\!\!\symsub{2}{x}\!=\!\left(\!\!\begin{array}{rrr} 1 & & \\ & -1 & \\ & & -1 \end{array}\!\!\right),
&&\!\!\symsub{\bar 6}{z}\!=\!\frac{1}{2}\!\left(\!\!\begin{array}{rrr} -1 & -\sqrt{3} & \\ \sqrt{3} & -1 & \\ & & -2 \end{array}\!\!\right).
\end{align*}

\autoref{eq:reactive} with $R=\symsub{2}{x}$ implies that $\Lambda_{ij}$ vanishes when $x$ appears an odd number of times in the indices $ij$. 
(That is, $\Lambda_{xj}=\Lambda_{ix}=0$ for $i,j=y,z$.) 
Similarly, $R=\symsub{m}{y}$ implies that $\Lambda_{ij}$ vanishes when $y$ appears an even number of times in the indices $ij$. 
(That is, $\Lambda_{ij}=0$ for $i,j=x,z$ and $\Lambda_{yy}=0$.) 
Consequently, only $\Lambda_{yz}$ and $\Lambda_{zy}$ are invariant under the symmetry operations $\symsub{2}{x}$ and $\symsub{m}{y}$. 
The operation $\symsub{\bar 6}{z}$ gives
\[
\Lambda_{yz}=-\tfrac{1}{2}\Lambda_{yz} \quad\text{and}\quad \Lambda_{zy}=-\tfrac{1}{2}\Lambda_{zy}
\]
for these elements. 
These relations can only hold for $\Lambda_{ij}=0$. 
Thus we conclude that $\Lambda_{ij}=0\,\,\forall\,\,i,j$. 

Repeating the analysis for $\Gamma_{ijk}$ with \autoref{eq:dissipative}, $R=\symsub{2}{x}$ implies that $\Gamma_{ijk}$ vanishes when $x$ appears an even number of times in the indices $ijk$, 
and $R=\symsub{m}{y}$ implies that $\Gamma_{ijk}$ vanishes when $y$ appears an odd number of times in the indices $ijk$. 
Consequently, only $\Gamma_{yyx}$, $\Gamma_{xzz}$, $\Gamma_{xxx}$, 
and the four other elements generated by freely permuting the indices $yyx$ and $xzz$ 
are invariant under the symmetry operations $\symsub{2}{x}$ and $\symsub{m}{y}$. 
The operation $\symsub{\bar 6}{z}$ gives
\[
\Gamma_{xzz}=-\tfrac{1}{2}\Gamma_{xzz}, \quad\Gamma_{zxz}=-\tfrac{1}{2}\Gamma_{zxz}, \quad\text{and}\quad\Gamma_{zzx}=-\tfrac{1}{2}\Gamma_{zzx},
\]
which implies that $\Gamma_{xzz}=\Gamma_{zxz}=\Gamma_{zzx}=0$. 
Furthermore,
\begin{align*}
&\Gamma_{yyx}=\tfrac{1}{8}(-\phantom{3}\Gamma_{yyx}+3\Gamma_{yxy}+3\Gamma_{xyy}-3\Gamma_{xxx}), \\
&\Gamma_{yxy}=\tfrac{1}{8}(+3\Gamma_{yyx}-\phantom{3}\Gamma_{yxy}+3\Gamma_{xyy}-3\Gamma_{xxx}), \\
&\Gamma_{xyy}=\tfrac{1}{8}(+3\Gamma_{yyx}+3\Gamma_{yxy}-\phantom{3}\Gamma_{xyy}-3\Gamma_{xxx}),
\end{align*}
and
\[
\Gamma_{xxx}=-\tfrac{1}{8}[\Gamma_{xxx}+3(\Gamma_{yyx}+\Gamma_{yxy}+\Gamma_{xyy})].
\]
Together these relations imply $\Gamma_{yyx}=\Gamma_{yxy}=\Gamma_{xyy}=-\Gamma_{xxx}$. 
We conclude that $\Gamma_{ijk}$ has four nonzero components, but only one free parameter, $\Gamma_{xxx}=\Gamma_{0}$. 
The effective field corresponding to the spin--orbit torque in \ce{Fe3GeTe2} is thus
\begin{equation}
\vec H_\text{SOT}=\Gamma_{0}[(m_x\,J_x-m_y\,J_y)\vec e_x-(m_y\,J_x+m_x\,J_y)\vec e_y]. \label{eq:torque}
\end{equation}

\section{Magnon spin and energy spectrum}
Using the result that the spin-orbit torque leads to a set of perpendicular in-plane easy and hard axes, as derived in the manuscript, we can write a model Hamiltonian in zero-external field,
\begin{align}
\nonumber\mathcal{H} = &-\frac{\varepsilon_J}{2\hbar^2}\sum_{\vec{r}}\sum_{\vec{\delta}} \vec{S}_{\vec{r}}\cdot\vec{S}_{\vec{r}+\vec{\delta}} -\frac{\varepsilon_z}{2\hbar^2}\sum_{\vec{r}}\left(S_{\vec{r},z}\right)^2 \\
&- \frac{\varepsilon_x}{2\hbar^2}\sum_{\vec{r}}\left[\left(S_{\vec{r},x}\right)^2-\left(S_{\vec{r},y}\right)^2\right] \, .
\label{eq:Hamiltonian}
\end{align}
Here we only consider nearest-neighbour exchange interaction between sites separated by $\vec{\delta}$, and only consider a current in the $x$-direction ($\varepsilon_x\propto \Gamma_0 J_x>0$), as the anisotropy behaves similarly (just with different axes) if we have a $y$-component of the current.

\subsection{Below the critical current}
\begin{figure}
  \includegraphics[width=.9\columnwidth]{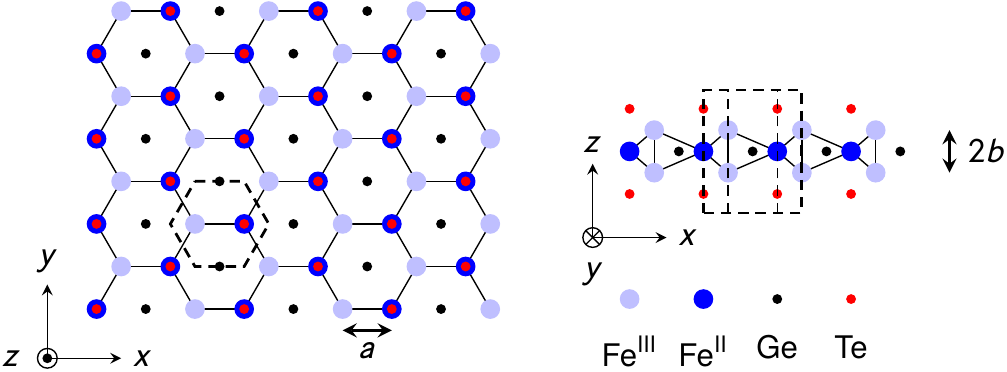}
  \caption{\label{fig:crystal}
	Crystal structure of monolayer \ce{Fe3GeTe2}. 
	All the drawn in-plane bindings are at a $120^\circ$ (in-plane) angle relative the neighboring bindings.
	Dashed lines denote the unit cell. 
	Left: view along $\vec e_z$; right: view along $\vec e_y$. 
	\ce{Fe^{III}} and \ce{Fe^{II}} represent the two inequivalent \ce{Fe} sites in oxidation states $+3$ and $+2$, respectively. 
	Redrawn after Ref.~\cite{Deng2018a}.
  }
\end{figure}

Below the critical current, the equilibrium configuration is along the $z$ axis.
We do a Holstein--Primakoff transformation of the spin operators, defined by
\begin{align}
S_{i,\nu,+} &= \hbar\sqrt{2s_\nu} a_{i,\nu}^\dagger\sqrt{1-\frac{a_{i,\nu}^\dagger a_{i,\nu}}{2s_\nu}} \approx \hbar\sqrt{2s_\nu} a_{i,\nu}^\dagger \, , \\
S_{i,\nu,-} &= \hbar\sqrt{2s_\nu} \sqrt{1-\frac{a_{i,\nu}^\dagger a_{i,\nu}}{2s_\nu}}a_{i,\nu} \approx \hbar\sqrt{2s_\nu} a_{i,\nu}\, , \\
S_{i,\nu,z} &= \hbar \left(a_{i,\nu}^\dagger a_{i,\nu}-s_\nu\right) \, ,
\label{eq:HP_Sz}
\end{align}
where $S_\pm = S_x\pm i S_y$, $i$ is the label of the unit cell, and $\nu=2,3_\pm$ indicates the sublattice of the \ce{Fe^{II}} and \ce{Fe^{III}} atoms, respectively, where sublattice $\nu=3_+$ ($\nu=3_-$) consists of the \ce{Fe^{III}} atoms located at $z=+b$ ($z=-b$).
We assume the nearest-neighbor exchange interaction is only between sublattice $\nu=2$ and $\nu=3_\pm$, and that there is no exchange interaction between the \ce{Fe^{III}} atoms.
Rewriting the Hamiltonian to the $S_\pm$ basis, we get
\begin{align}
\nonumber\mathcal{H} = &-\frac{\varepsilon_J}{2\hbar^2}\sum_{\vec{r},\vec{\delta}} \left[\frac{1}{2}\left(S_{\vec{r},+}S_{\vec{r}+\vec{\delta},-}+S_{\vec{r},-}S_{\vec{r}+\vec{\delta},+}\right)+S_{\vec{r},z}S_{\vec{r}+\vec{\delta},z}\right] \\
&-\frac{\varepsilon_z}{2\hbar^2}\sum_{\vec{r}}\left(S_{\vec{r},z}\right)^2 - \frac{\varepsilon_x}{4\hbar^2}\sum_{\vec{r}}\sum_{m=\pm}\left(S_{\vec{r},m}\right)^2 \, .
\end{align}
Inserting the Holstein--Primakoff transformation, keeping terms to second order in the magnon operators, we get
\begin{align}
\nonumber\mathcal{H} = &-\varepsilon_J\sum_{i}\sum_{\vec{r}_j=\vec{r}_i+\vec{\delta}}\sum_{\nu=3_\pm}\Big[\sqrt{s_{2} s_{\nu}}\left(a_{i,{2}}^\dagger a_{j,\nu}+a_{j,\nu}^\dagger a_{i,2}\right) \\
\nonumber &-s_{\nu}a_{i,2}^\dagger a_{i,2} - s_{2} a_{j,\nu}^\dagger a_{j,\nu} \Big] + \varepsilon_z\sum_{i,\nu}\ s_\nu a_{i,\nu}^\dagger a_{i,\nu} \\
&- \frac{\varepsilon_x}{2}\sum_{i,\nu} s_\nu (a_{i,\nu}^\dagger a_{i,\nu}^\dagger+a_{i,\nu} a_{i,\nu}) \, ,
\end{align}
disregarding any constant terms, where $\vec{r}_j$ is the position of the nearest-neighbor atom of the atom located in unit cell $i$ and sublattice $\nu=2$.
Next we perform a Fourier transform to momentum space, defined by
\begin{align}
a_{i,\nu} = \frac{1}{\sqrt{N}}\sum_{\vec{k}} a_{\vec{k},\nu}e^{-i\vec{k}\cdot\vec{r}_{i,\nu}} \, , \quad a_{i,\nu}^\dagger = \frac{1}{\sqrt{N}}\sum_{\vec{k}} a_{\vec{k},\nu}^\dagger e^{i\vec{k}\cdot\vec{r}_{i,\nu}} \, ,
\end{align}
with $N$ being the number of unit cells, and $\vec{k}$ the wave vector running over the first Brillouin zone.  
The Hamiltonian then becomes (disregarding any constant terms)
\begin{align}
&\nonumber\mathcal{H} = \sum_{\vec{k}}\sum_{z=\pm}\varepsilon_J\Big[3s_{3}a_{\vec{k},2}^\dagger a_{\vec{k},2} + 3 s_{2} a_{\vec{k},3_z}^\dagger a_{\vec{k},3_z} \\
\nonumber &-\sqrt{s_{2} s_{3}}\left(\gamma_{-\vec{k}}^z a_{\vec{k},2}^\dagger a_{\vec{k},3_z}+\gamma_{\vec{k}}^z a_{\vec{k},3_z}^\dagger a_{\vec{k},{2}}\right)\Big] +\varepsilon_z \sum_{\vec{k},\nu}s_\nu a_{\vec{k},\nu}^\dagger a_{\vec{k},\nu} \\
&-\frac{\varepsilon_x}{2} \sum_{\vec{k},\nu} s_\nu \left(a_{\vec{k},\nu}^\dagger a_{-\vec{k},\nu}^\dagger+a_{\vec{k},\nu} a_{-\vec{k},\nu}\right) \, .
\end{align}
Here we have introduced the structure factor
\begin{align}
\gamma_{\vec{k}} = \sum_{\vec{\delta}} e^{i\vec{k}\cdot\vec{\delta}} \, ,
\end{align}
which becomes
\begin{align}
\gamma_{\vec{k}}^\pm = e^{\pm i k_z b}\left[e^{-i k_x a}+2e^{i k_x a/2}\cos\left(\frac{\sqrt{3}}{2}k_y a\right)\right]
\end{align}
for Fe$_3$GeTe$_2$ between the $\nu=2$ and $\nu=3_\pm$ sublattices, as can be seen from \autoref{fig:crystal}. Here $a$ is the in-plane lattice constant between the \ce{Fe^{II}} and \ce{Fe^{III}} atoms, and $2b$ the separation between two \ce{Fe^{III}} atoms in the $z$ direction.
We have also used that there are three nearest neighbors in each sublattice.
We can write the Hamiltonian on the form
\begin{align}
\nonumber &\mathcal{H} = \sum_{\vec{k}}\Bigg( \frac{A}{2}a_{\vec{k},2}^\dagger a_{\vec{k},2}+\frac{B}{2}a_{\vec{k},3_-}^\dagger a_{\vec{k},3_-}+\frac{B}{2}a_{\vec{k},3_+}^\dagger a_{\vec{k},3_+}\\
&+C_{\vec{k}} a_{\vec{k},2}^\dagger a_{\vec{k},3_-} +D_{\vec{k}} a_{\vec{k},2}^\dagger a_{\vec{k},3_+} +\sum_\nu E_\nu a_{\vec{k},\nu} a_{-\vec{k},\nu} \Bigg) + \text{H.c.}
\label{eq:GeneralHamiltonian}
\end{align}
The coefficients $A$, $B$, $C_{\vec{k}}$, $D_{\vec{k}}$, and $E_\nu$ are given in \autoref{tab:coefficients}.

We now have to diagonalize the Hamiltonian.
This can be done by a six-dimensional Bogoliubov transformation, defined by the matrix $\underline{B}_6$
\begin{align}
&\nonumber\vec{\alpha}_{\vec{\kappa}} =
\begin{pmatrix}
\alpha_{\vec{\kappa},\mathrm{I}} \\ \alpha_{\vec{\kappa},\mathrm{II}} \\ \alpha_{\vec{\kappa},\mathrm{III}} \\\alpha_{-\vec{\kappa},\mathrm{I}}^\dagger \\ \alpha_{-\vec{\kappa},\mathrm{II}}^\dagger \\ \alpha_{-\vec{\kappa},\mathrm{III}}^\dagger
\end{pmatrix}
=
\underline{B}_6
\begin{pmatrix}
a_{\vec{\kappa},2} \\ a_{\vec{\kappa},3_-} \\ a_{\vec{\kappa},3_+} \\ a_{-\vec{\kappa},2}^\dagger \\ a_{-\vec{\kappa},3_-}^\dagger \\  a_{-\vec{\kappa},3_+}^\dagger
\end{pmatrix} 
\equiv \underline{B}_6
\vec{a}_{\vec{\kappa}} \\
&= \sum_\nu
\begin{pmatrix}
u_{\mathrm{I},\nu} & v_{\mathrm{I},\nu} \\
u_{\mathrm{\mathrm{I}I},\nu} & v_{\mathrm{\mathrm{I}I},\nu} \\
u_{\mathrm{\mathrm{\mathrm{I}I}I},\nu} & v_{\mathrm{\mathrm{\mathrm{I}I}I},\nu} \\
\tilde{v}_{\mathrm{I},\nu}^* & \tilde{u}_{\mathrm{I},\nu}^* \\
\tilde{v}_{\mathrm{\mathrm{I}I},\nu}^* & \tilde{u}_{\mathrm{\mathrm{I}I},\nu}^* \\
\tilde{v}_{\mathrm{\mathrm{\mathrm{I}I}I},\nu}^* & \tilde{u}_{\mathrm{\mathrm{\mathrm{I}I}I},\nu}^*
\end{pmatrix}
\begin{pmatrix}
a_{\vec{\kappa},\nu} \\
a_{-\vec{\kappa},\nu}^\dagger
\end{pmatrix}\, ,
\label{eq:Bogoliubov6D}
\end{align}
where $\vec{\kappa}$ now only runs over half the vector space of $\vec{k}$, so that the Hamiltonian can be written as
\begin{align}
\mathcal{H}=\sum_{\vec{\kappa},\mu}\left(\varepsilon_{\vec{\kappa},\mu}\alpha_{\vec{\kappa},\mu}^\dagger\alpha_{\vec{\kappa},\mu}+\varepsilon_{-\vec{\kappa},\mu}\alpha_{-\vec{\kappa},\mu}^\dagger\alpha_{-\vec{\kappa},\mu}\right) \, .
\end{align}
The Bogoliubov coefficients with a tilde, e.g. $\tilde{v}_{\mathrm{I},2}$, are evaluated at $-\vec{\kappa}$ while the coefficients without tilde are evaluated at $\vec{\kappa}$. 
To diagonalize the Hamiltonian we impose bosonic commutation relations ($[\alpha_{\vec{\kappa},\mu},\alpha_{\vec{\kappa'},\mu'}^\dagger ]=\delta_{\vec{\kappa},\vec{\kappa'}}\delta_{\mu,\mu'}$) as well as the relation $\left[\alpha_{\vec{\kappa},\mu},\mathcal{H}\right]=\varepsilon_{\vec{\kappa},\mu}\alpha_{\vec{\kappa},\mu}$.
The bosonic commutation relation leads to the constraint
\begin{align}
\left[\vec{\alpha}_{\vec{\kappa}},\vec{\alpha}_{\vec{\kappa}}^\dagger \right] = \underline{B}_6\left[\vec{a}_{\vec{\kappa}},\vec{a}_{\vec{\kappa}}^\dagger \right]\underline{B}_6^\dagger = \underline{B}_6\underline{Y}\underline{B}_6^\dagger = \underline{Y} \, ,
\label{eq:BosonicCommRels}
\end{align}
where we have introduced the matrix
\begin{align}
\underline{Y}=
\diag(1,1,1,-1,-1,-1)\,.
\end{align}
The relation in \autoref{eq:BosonicCommRels} requires the normalization
\begin{equation}
\sum_\nu \left(\abs{u_{\mu\nu}}^2-\abs{v_{\mu\nu}}^2\right) = 1 \, .
\label{eq:BogoliubovNormalization}
\end{equation}
The relation from the commutation with the Hamiltonian leads to the eigenvalue problem
\begin{align}
\begin{pmatrix}
A & C_{\vec{\kappa}}^* & D_{\vec{\kappa}}^* & -2E_2 & & \\
C_{\vec{\kappa}} & B & & & -2E_{3_-} & \\
D_{\vec{\kappa}} & & B & & & -2E_{3_+} \\
2E_2 & & & -A & -C_{-\vec{\kappa}} & -D_{-\vec{\kappa}} \\
& 2E_{3_-} & & -C_{-\vec{\kappa}}^* & -B & \\
& & 2E_{3_+} & -D_{-\vec{\kappa}}^* & & -B
\end{pmatrix}
\vec{e}_\mu
=
\varepsilon_{\vec{\kappa},\mu}
\vec{e}_\mu \, ,
\end{align}
where $\vec{e}_\mu={(u_{\mu,2}, u_{\mu,3_-}, u_{\mu,3_+}, v_{\mu,2}, v_{\mu,3_-}, v_{\mu,3_+})^\mathrm{T}}$.
We note that $C_{-\vec{\kappa}} = C_{\vec{\kappa}}^*$ and $D_{-\vec{\kappa}} = D_{\vec{\kappa}}^*$, and all other elements in the matrix are real and independent of $\vec{\kappa}$. Consequently, we therefore have that $\tilde{u}_{\mu,\nu} = u_{\mu,\nu}^*$ and $\tilde{v}_{\mu,\nu} = v_{\mu,\nu}^*$.
We also have that $\varepsilon_{\vec{\kappa},\mu}^* = \varepsilon_{-\vec{\kappa},\mu}$, and as $\varepsilon_{\vec{\kappa},\mu}$ is a real quantity, we therefore also have $\varepsilon_{-\vec{\kappa},\mu}=\varepsilon_{\vec{\kappa},\mu}$.

\begin{table}[tpb]
\centering
\caption{The coefficients for the Fourier transformed Hamiltonian in \autoref{eq:GeneralHamiltonian} below and above the critical current $\abs{J_c}$.}
\begin{tabular}{c c c}
\hline \hline
Coefficient & $\quad\quad\quad\abs{J}<\abs{J_c}\quad\quad\quad$ & $\abs{J}>\abs{J_c}$  \\
\hline
$A$ & $6s_3\varepsilon_J+s_2\varepsilon_z$ & $6s_3\varepsilon_J+\frac{1}{2}s_2(3\varepsilon_x-\varepsilon_z)$\\
$B$ & $3s_2\varepsilon_J+s_3\varepsilon_z$ & $3s_2\varepsilon_J+\frac{1}{2}s_3(3\varepsilon_x-\varepsilon_z)$\\
$C_{\vec{k}}$ & $-\sqrt{s_2s_3}\gamma_{-\vec{k}}^-\varepsilon_J$ & $-\sqrt{s_2s_3}\gamma_{-\vec{k}}^-\varepsilon_J$ \\
$D_{\vec{k}}$ & $-\sqrt{s_2s_3}\gamma_{-\vec{k}}^+\varepsilon_J$ & $-\sqrt{s_2s_3}\gamma_{-\vec{k}}^+\varepsilon_J$\\
$E_\nu$ & $-\frac{1}{2}s_\nu\varepsilon_x$ & $-\frac{1}{4}s_\nu(\varepsilon_x+\varepsilon_z)$ \\
\hline \hline
\end{tabular}
\label{tab:coefficients}
\end{table}

In addition to finding the energy of the eigenmagnons, we also wish to determine their spin, as these are not integer due to squeezing from the SOT-induced anisotropy~\cite{Kamra:prb:2017}. Using \autoref{eq:Bogoliubov6D} and \autoref{eq:BosonicCommRels} we see that $\vec{a}_{\vec{\kappa}} = {\underline{B}_6^{-1}\vec{\alpha}_{\vec{\kappa}}} = {\underline{Y}\underline{B}_6^\dagger\underline{Y}^{-1}}$. This can be written explicitly as
\begin{align}
a_{\vec{\kappa},\nu} &= \sum_\mu \left(u_{\mu,\nu}\alpha_{\vec{\kappa},\mu}-v_{\mu,\nu}\alpha_{-\vec{\kappa},\mu}^\dagger\right) \, ,\\
a_{\vec{\kappa},\nu}^\dagger &= \sum_\mu \left(u_{\mu,\nu}^*\alpha_{\vec{\kappa},\mu}^\dagger -v_{\mu,\nu}^*\alpha_{-\vec{\kappa},\mu}\right) \, . 
\end{align}
Together with the fact that non-diagonal expectation values of the product of two eigenmagnon operators vanish, we see from \autoref{eq:HP_Sz} and \autoref{eq:BogoliubovNormalization} that
\begin{align}
\nonumber\sum_{i,\nu}\langle S_{i,\nu,z}\rangle &= \sum_{\vec{\kappa},\mu} \hbar\sum_\nu\left(\abs{u_{\mu,\nu}}^2+\abs{v_{\mu,\nu}}^2\right)\sum_{m=\pm}\langle\alpha_{m\vec{\kappa},\mu}^\dagger \alpha_{m\vec{\kappa},\mu}\rangle  \\
&= \sum_{\vec{k},\mu} \hbar\left(1+2\sum_\nu\abs{v_{\mu,\nu}}^2\right)\langle\alpha_{\vec{k},\mu}^\dagger \alpha_{\vec{k},\mu}\rangle \, ,
\end{align}
where we have disregarded all constant terms.
We can then see that the eigenmagnon spin contribution is
\begin{align}
S_{\vec{k},\mu} = \hbar\left(1+2\sum_\nu\abs{v_{\mu,\nu}}^2\right) \, .
\end{align}

\subsection{Above the critical current}
Above the critical current, the lowest energy configuration of the spins is along the $x$ axis.
We therefore have to change the Holstein--Primakoff transformation to reflect this, with the following transformation:
\begin{align}
\tilde{S}_{i,\nu,+} &= \hbar\sqrt{2s_\nu} a_{i,\nu}^\dagger\sqrt{1-\frac{a_{i,\nu}^\dagger a_{i,\nu}}{2s_\nu}} \approx \hbar\sqrt{2s_\nu} a_{i,\nu}^\dagger \, , \\
\tilde{S}_{i,\nu,-} &= \hbar\sqrt{2s_\nu} \sqrt{1-\frac{a_{i,\nu}^\dagger a_{i,\nu}}{2s_\nu}}a_{i,\nu} \approx \hbar\sqrt{2s_\nu} a_{i,\nu}\, , \\
\tilde{S}_{i,\nu,x} &= \hbar \left(a_{i,\nu}^\dagger a_{i,\nu}-s_\nu\right) \, ,
\end{align}
with $\tilde{S}_\pm = -\tilde{S}_z\pm i\tilde{S}_y$. Using this transformation in the Hamiltonian in \autoref{eq:Hamiltonian}, we get
\begin{align}
\nonumber\mathcal{H} = &-\varepsilon_J\sum_{i}\sum_{\vec{r}_j=\vec{r}_i+\vec{\delta}}\sum_{\nu=3_\pm}\Big[\sqrt{s_{2} s_{\nu}}\left(a_{i,{2}}^\dagger a_{j,\nu}+a_{j,\nu}^\dagger a_{i,2}\right) \\
\nonumber &-s_{\nu}a_{i,2}^\dagger a_{i,2} - s_{2} a_{j,\nu}^\dagger a_{j,\nu} \Big]\\
\nonumber & -\frac{\varepsilon_z}{4}\sum_{i,\nu} s_\nu \left( a_{i,\nu}^\dagger a_{i,\nu}^\dagger+2a_{i,\nu}^\dagger a_{i,\nu} + a_{i,\nu}a_{i,\nu}\right)  \\
&+ \frac{\varepsilon_x}{4}\sum_{i,\nu} s_\nu \left(6a_{i,\nu}^\dagger a_{i,\nu}-a_{i,\nu}^\dagger a_{i,\nu}^\dagger-a_{i,\nu} a_{i,\nu}\right) \, .
\end{align}
We again do a Fourier transformation as before, and find the Hamiltonian to be on the form (again disregarding any constant terms)
\begin{align}
\nonumber\mathcal{H} = &+\sum_{\vec{k}}\sum_{z=\pm}\varepsilon_J\Big[3s_{3}a_{\vec{k},2}^\dagger a_{\vec{k},2} + 3 s_{2} a_{\vec{k},3_z}^\dagger a_{\vec{k},3_z} \\
\nonumber &-\sqrt{s_{2} s_{3}}\left(\gamma_{-\vec{k}}^z a_{\vec{k},2}^\dagger a_{\vec{k},3_z}+\gamma_{\vec{k}}^z a_{\vec{k},3_z}^\dagger a_{\vec{k},{2}}\right)\Big] \\
\nonumber &-\frac{\varepsilon_z}{4}\sum_{\vec{k},\nu} s_\nu \left(2a_{\vec{k},\nu}^\dagger a_{\vec{k},\nu} + a_{\vec{k},\nu}^\dagger a_{-\vec{k},\nu}^\dagger + a_{\vec{k},\nu}a_{-\vec{k},\nu}\right) \\
&+ \frac{\varepsilon_x}{4}\sum_{\vec{k},\nu} s_\nu \left(6a_{\vec{k},\nu}^\dagger a_{\vec{k},\nu}-a_{\vec{k},\nu}^\dagger a_{-\vec{k},\nu}^\dagger-a_{\vec{k},\nu} a_{-\vec{k},\nu}\right) \, .
\end{align}
From this expression we can read off the coefficients in \autoref{eq:GeneralHamiltonian}, and use the results in the previous subsection for the case below the critical current to determine the energy and spin of the eigenmagnons.
The coefficients in \autoref{eq:GeneralHamiltonian} are given in \autoref{tab:coefficients} both above and below the critical current.


%